\documentclass[aps,prl, reprint,amsmath,amssymb,longbibliography]{revtex4-2}

\usepackage{graphicx,epsfig,epsf,color,hhline,import}% Include figure files
\usepackage{subcaption}
\usepackage{dcolumn}% Align table columns on decimal point
\usepackage{bm}% bold math
\usepackage{tensor,pbox}
\usepackage{tikz}
\usepackage{epstopdf}
\usepackage{amsthm}
\usepackage{bbold}
\usepackage[ruled,vlined]{algorithm2e}

\usepackage{changes}
\usepackage{cancel}
\usepackage{color}
\usepackage{url}
\usepackage[hidelinks]{hyperref}
\usepackage{ytableau}
\usepackage{enumerate}
\usepackage{mathdots}
\usepackage{braket}
\usepackage{MnSymbol}
\usepackage{mathtools}

\newcommand{\one}{\mathbb{1}}
\newcommand{\tr}{{\rm tr}}

\newcommand{\p}[2]{p\left(#1|#2\right)}
\newcommand{\bt}{{\boldsymbol{\theta}}}

\newcommand{\pnorm}[2]{\ensuremath{ \left\Vert  #1  \right\Vert_{#2} }}

\newcommand{\e}[1]{\ensuremath{\operatorname{e}^{#1}}}

\definecolor{bostonuniversityred}{rgb}{0.8, 0.0, 0.0}

\begin{abstract}
  Quantum metrology promises unprecedented measurement precision but suffers in practice from the limited availability of resources such as the number of probes, their coherence time, or non-classical quantum states. The adaptive Bayesian approach to parameter estimation allows for an efficient use of resources thanks to adaptive experiment design. For its practical success fast numerical solutions for the
  Bayesian update and the adaptive experiment design are crucial.
  Here we show that neural networks can be trained to become fast and strong experiment-design heuristics using a combination of an evolutionary strategy and reinforcement learning. Neural-network heuristics are shown to outperform
  established heuristics for the technologically important example of frequency estimation
  of a qubit that suffers from dephasing. Our method of creating neural-network heuristics is very general and complements the well-studied sequential Monte Carlo method for Bayesian updates to form a complete framework for adaptive Bayesian quantum estimation.   
\end{abstract}
\begin{document}
\title{Neural-Network Heuristics for Adaptive Bayesian Quantum Estimation}
\author{Lukas J.~Fiderer$^{1,2}$, Jonas Schuff$^{1,3}$, Daniel Braun$^1$}
\affiliation{$^1$Eberhard-Karls-Universit\"at T\"ubingen, Institut f\"ur Theoretische Physik, 72076 T\"ubingen, Germany\\
$^2$School of Mathematical Sciences, The University of Nottingham, University Park, Nottingham NG7 2RD, United Kingdom\\
$^3$Department of Materials, University of Oxford, Parks Road, Oxford OX1 3PH, United Kingdom}
\maketitle
In quantum metrology we aim to design quantum experiments such that  one or multiple parameters can be estimated from the measurement outcomes.  
Experiment design can involve the preparation of initial states, controlling the dynamics, or choosing measurements for readout. The estimation of parameters is a problem of statistical inference, and the most common approaches to tackle it are the frequentist and the Bayesian one.

In the frequentist approach, experiments are typically repeated
several times which allows one to estimate the parameters from the statistics of measurement outcomes using, for example, maximum likelihood estimation. The problem of experiment design is often addressed with the Cram\'er--Rao bound formalism \cite{helstrom_quantum_1976, braunstein_statistical_1994} by maximizing the quantum Fisher information with respect to experiment designs \cite{fiderer2019maximal}.

The Bayesian approach, on the other hand, relies on updating the current knowledge of the parameters after each experiment using Bayes' law. Examples for Bayesian quantum estimation involve state and process tomography \cite{buzek1998reconstruction, blumekohout2010optimal,huszar2012adaptive,granade2012robust,granade2016practical, granade2017practical}, and phase and frequency estimation \cite{ferrie2012how,wiebe2014hamiltonian,wiebe2016efficient,friis2017flexible} with various experimental realizations \cite{kravtsov2013experimental,struchalin2016experimental,chapman2016experimental,hou2019experimental,wang2017experimental,paesani2017experimental,lumino2018experimental,santagati2019magnetic}.  The Bayesian approach is particularly suitable for adaptive experiment design: experiments can be optimized depending on the current knowledge of the parameters and the available resources. While adaptivity can enhance the precision and save time and other resources compared with non-adaptive (frequentist) approaches \cite{chaloner1995bayesian, sergeevich2011characterization}, it involves a computational challenge:
The Bayesian update and the consecutive optimization of the experiment design are both analytically intractable (with rare exceptions under idealized conditions \cite{childs2000quantum, ferrie2012how}). In view of the short timescale of quantum experiments, slow numerical computation of the  Bayesian update and the experiment design can drastically increase the total time consumed.
In order to approximate the Bayesian update efficiently, a framework based on a sequential Monte Carlo (SMC) algorithm has been developed \cite{liu2001combined, huszar2012adaptive,granade2012robust,granade2017qinfer, qinfer}. This framework, however, does not solve the problem of adaptive experiment design, which represents a second computational step.

\begin{figure*}
	\centering
	\begin{subfigure}{.34\textwidth}
	\centering
	\includegraphics[width=0.9\linewidth]{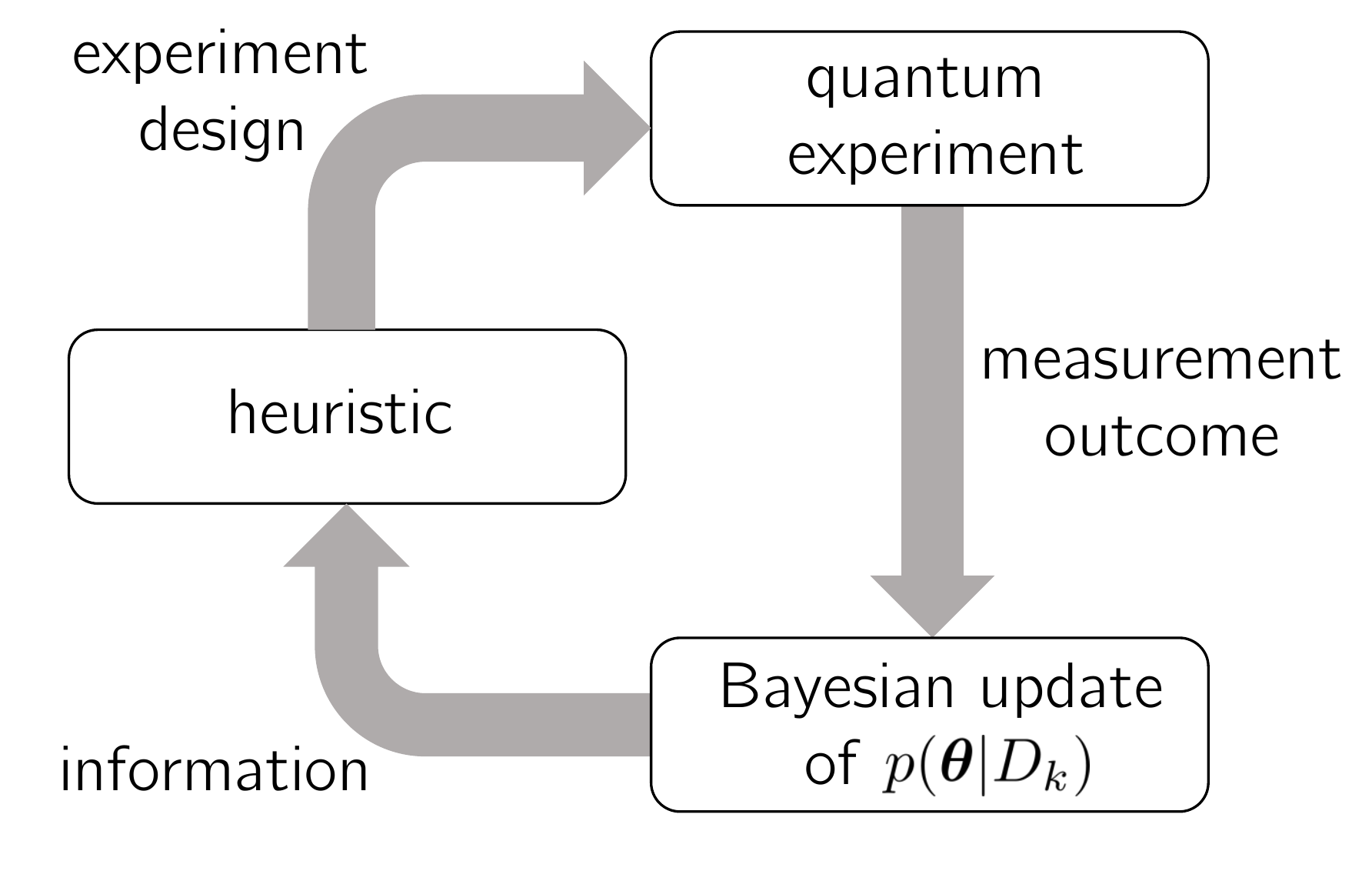}\\
	(a) Adaptive Bayesian quantum estimation.\\
	\vspace{1cm}
	\includegraphics[width=0.99\linewidth]{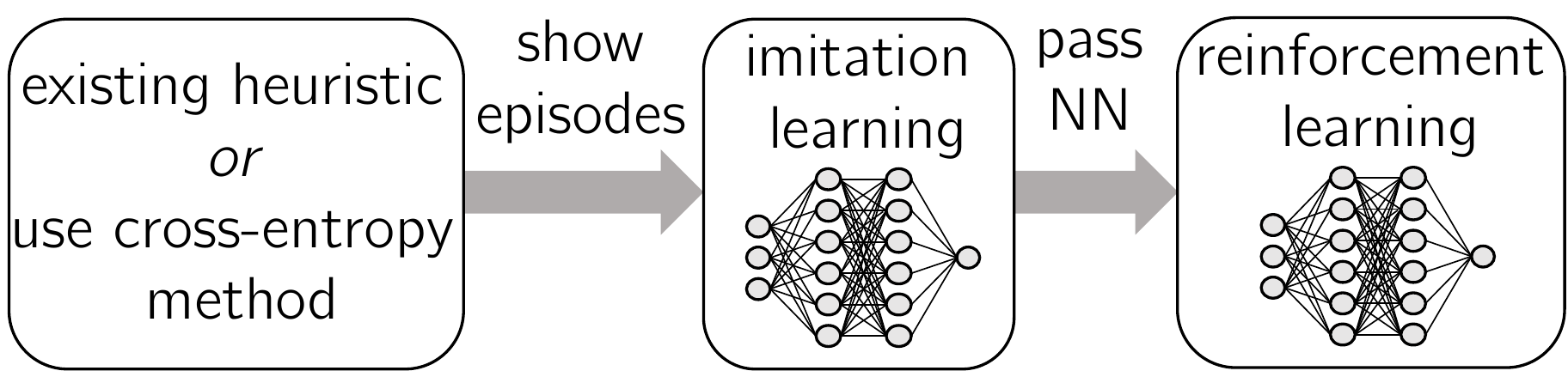}\\
	(b) Training procedure.
	\end{subfigure}
	\begin{subfigure}{.32\textwidth}
	\centering
	\includegraphics[width=0.99\linewidth]{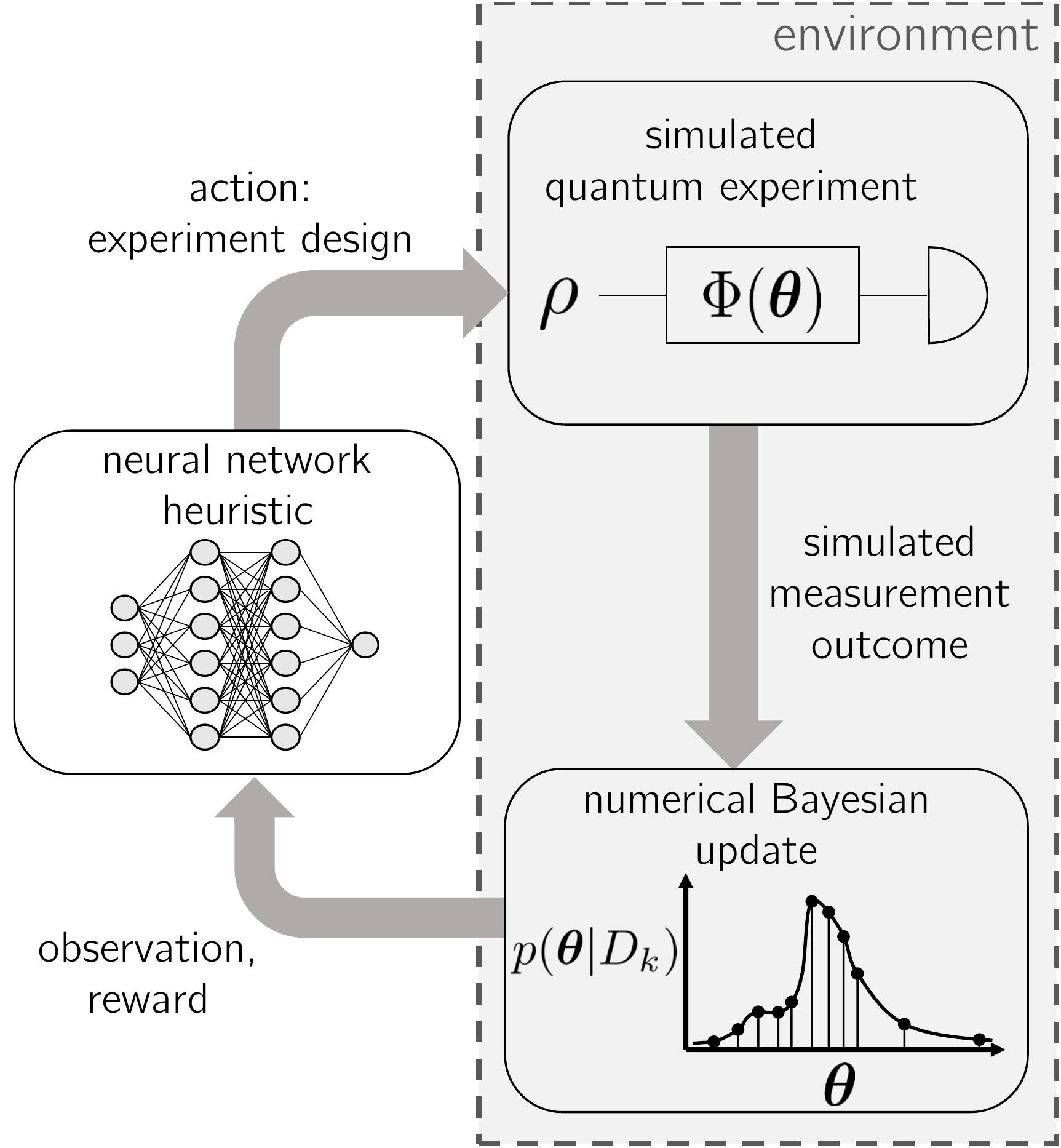}\\
	(c) Reinforcement learning setup.
	\end{subfigure}
	\begin{subfigure}{.32\textwidth}
	\centering
	\includegraphics[width=0.99\linewidth]{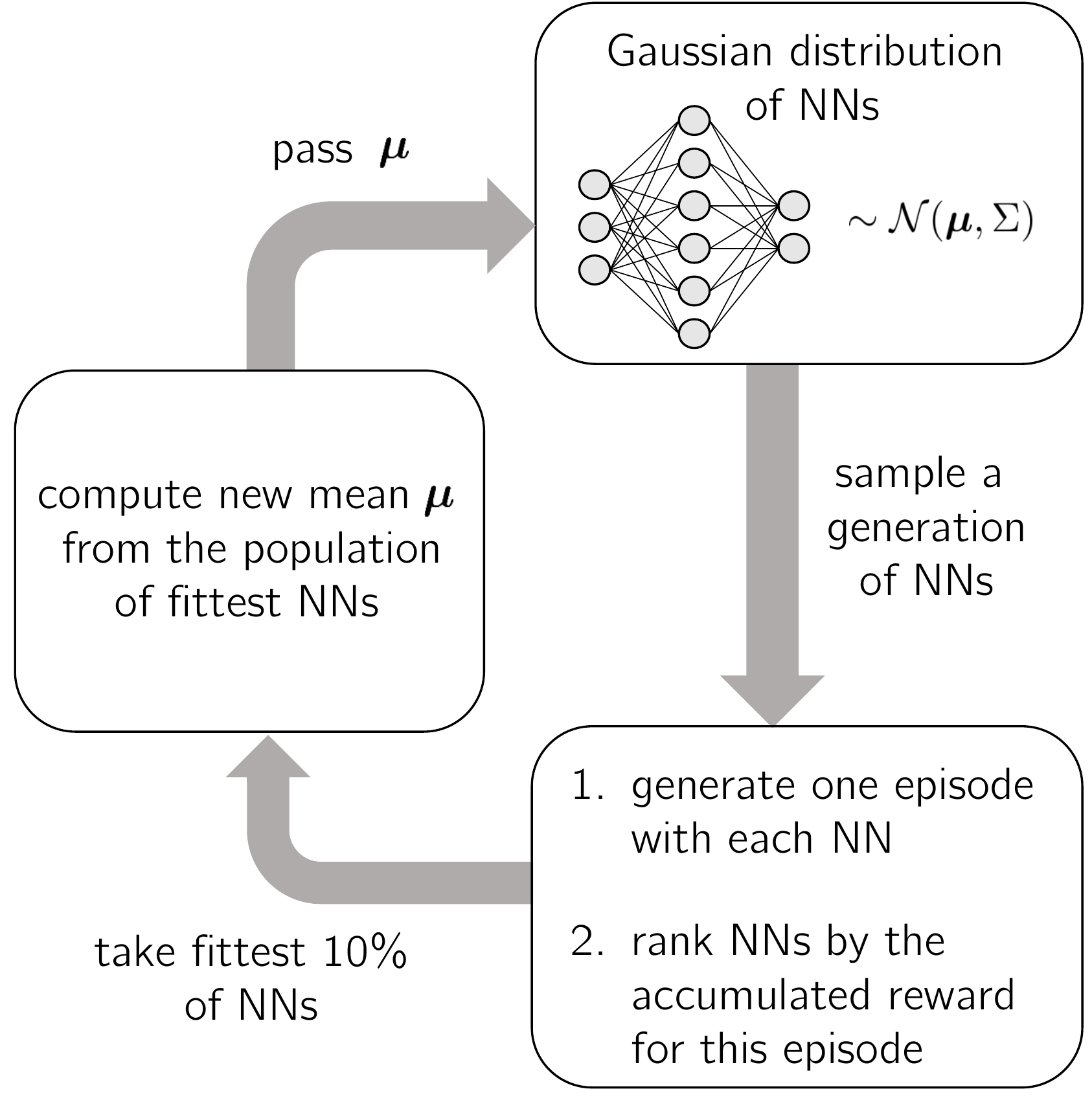}\\
	(d) Cross-entropy method.
	\end{subfigure}
	\caption{Panel (a) shows the setup for adaptive Bayesian quantum estimation. First, prior information (e.g., about the prior $p(\bt)$ and the available resources) is passed to the heuristic which returns an experiment design $e_1$. From the quantum experiment, we obtain a measurement outcome $d_1$ which is used to numerically compute the Bayesian update of $p(\bt)$. Then, it continues in the same manner and this procedure is repeated until some exit condition is fulfilled. Panel (b) depicts the training procedure of the neural networks (NNs). First, the NN learns to imitate the behavior of a known heuristic (which may have been found using the cross-entropy method). Then, we use this NN as a starting point for reinforcement learning (RL). Panel (c) shows the RL setup for generating training data. An agent (depicted as a NN) interacts with a RL environment. The RL environment defines the estimation problem. It consists of a simulated quantum experiment [depicted with initial state $\rho$ and parameter-dependent quantum channel $\Phi(\bt)$] and a numerical Bayesian update. It also stores information about available resources for experiment design. Panel (d) shows the evolutionary principle of a simple cross-entropy method. NNs are treated as vectors of parameters sampled from a Gaussian distribution $\mathcal{N}(\boldsymbol{\mu}, \Sigma)$ with constant covariance. At the beginning, the NNs are initialized randomly.}\label{fig:schemes}
\end{figure*}

In practice, one has to rely on so-called experiment-design heuristics, i.e., fast-to-evaluate functions which take  available information as input and return an experiment design as output, see Fig.~\ref{fig:schemes}(a). If the output depends
on the available information, we speak of an adaptive heuristic. 
So far, adaptive experiment-design heuristics for Bayesian estimation have been found  mostly manually, typically motivated by analytic arguments derived for idealized conditions concerning the experimental model and the available resources \cite{ferrie2012how, wiebe2014hamiltonian, stenberg2014efficient, granade2017practical, lumino2018experimental}. In one case a manually found heuristic \cite{stenberg2014efficient} has been fine-tuned offline using a particle swarm algorithm \cite{stenberg2016characterization}. Another approach is to  optimize experiment designs between the experiments with respect to a restricted set of experiment designs in order to keep the problem numerically tractable \cite{hincks2018hamiltonian}.	
Apart from Bayesian inference, particle swarm and differential evolution algorithms have been used to search a certain class of experiment-design heuristics (represented by binary decision trees) in order to optimize the scaling of the uncertainty in  phase estimation with the number of entangled photons in an interferometer \cite{hentschel2010machine,hentschel2011efficient,lovett2013differential,lumino2018experimental}.
A general numerical framework for finding adaptive experiment-design heuristics for Bayesian quantum estimation is missing so far. 

We consider an approach to experiment-design heuristics which uses reinforcement learning (RL). Recently, RL has been used with great success to create programs that play chess and Go better than any other program or human \cite{silver2018general}. RL has also been used in quantum physics \cite{palittapongarnpim2017learning,fosel2018reinforcement, dunjko2018machine,schafer2020differentiable} and, in particular, in quantum metrology for improving the dynamics of quantum sensors \cite{xu2019transferable, schuff2020improving}. Problems in quantum metrology have also been tackled with other machine-learning methods using neural-networks: supervised learning  has been used for calibrating quantum sensors \cite{cimini2019calibration}, unsupervised learning has been proposed for retrieving Hamiltonian parameters from experimental data \cite{gentile2020learning}, and a Bayes classifier has been used for the identification of light sources \cite{you2019identification}.

Here, we propose to use neural networks (NNs) as experiment-design heuristics. We provide a general method based on a combination of an evolutionary strategy with RL for creating such NN experiment-design heuristics. This method builds upon and complements the SMC framework for approximate Bayesian updates \cite{liu2001combined,huszar2012adaptive,granade2012robust, granade2017qinfer, qinfer}. It is general in the sense that (i) it can be easily adjusted to all kinds of estimation problems, (ii) it uses a very general ansatz for the experiment-design heuristic in form of a NN, and (iii) it 
creates a 
NN experiment-design heuristic
that can take
into account not only knowledge of the parameters to be estimated but also additional information, for instance, about available resources. Further, our method numerically approaches global optimization in the sense that it tries to find heuristics which are optimal for all future experiments as opposed to local (greedy) strategies which always optimize only the next experiment. The trained neural networks represent ready-to-use experiment-design heuristics and we envisage their application in Bayesian quantum sensors in combination with the SMC framework for approximate Bayesian updates \cite{liu2001combined,huszar2012adaptive,granade2012robust, granade2017qinfer, qinfer}.

\section{Bayes risk}

Let $\bt=(\theta_1,\dotsc, \theta_d)\in \Theta$ be a vector of parameters we want to estimate, and we assume that $\Theta$ restricts each $\theta_j$ to a finite interval. The prior $p(\bt)$ is a probability distribution on $\Theta$ which represents our knowledge prior to the first measurement, and we imagine that prior to each Bayesian estimation $\bt$ is sampled from $p(\bt)$. After each experiment, our knowledge of $\bt$ is updated with Bayes' law (see Appendix A). 
Let $\p{\bt}{D_k, E_k}$ represent this updated knowledge of $\bt$ after the $k$th measurement, where $D_k=(d_1,\dotsc, d_k)$ denotes the measurement outcomes from a sequence of $k$ experiments $E_k=(e_1,\dotsc, e_k)$.
In the following, we omit the dependence on experiment designs for the sake of clarity, e.g., we write $\p{\bt}{D_k}$ instead of $\p{\bt}{D_k, E_k}$.

An experiment-design heuristic $h$ is a function which maps available information, for instance about $\bt$ or about available resources, to an experiment design for the next experiment, see Fig.~\ref{fig:schemes} (a). The idea is to consult the experiment-design heuristic prior to each experiment and design the experiment accordingly. Imagine that the available resources for one Bayesian estimation are such that we can make $k$ experiments. Then, we aim to choose an experiment-design heuristic $h$ which minimizes the expected traced covariance over $p(\bt|D_k)$:
\begin{align}
r\left[h | p(\bt)\right] = \operatorname{E}_{D_k}\left(\tr\left[\operatorname{Cov}_{\bt|D_k}\left(\bt\right)\right]\right), \label{eq:Bayes}
\end{align}
where $\operatorname{Cov}_{\bt|D_k}=\operatorname{E}_{\bt|D_k}\left(\bt\bt^\text{T}\right)-\operatorname{E}_{\bt|D_k}\left(\bt\right)\operatorname{E}_{\bt|D_k}\left(\bt\right)^\text{T}$ is the covariance, and
the notation  $\operatorname{E}_{a|b}(c)$ denotes the expected value of $c$ with respect to $a \in A$ distributed as $p(a|b)$, $\operatorname{E}_{a|b}(c)=\int_A \text{d}a\, p(a|b) c$. If $a$ takes discrete values the integral becomes a sum, and if $a$ is distributed as $p(a)$ we write $\operatorname{E}_{a}(c)$ instead of $\operatorname{E}_{a|b}(c)$.

In Appendix B we show that $r\left[h | p(\bt)\right]$ corresponds to the Bayes risk for a loss function $L\left[\hat{\bt}_k(D_k)| \bt\right]=\pnorm{\hat{\bt}_k(D_k)-\bt}{2}^2$ with the Bayes estimator $\hat{\bt}_k$ after $k$ measurements given by $\hat{\bt}_k (D_k)= \operatorname{E}_{\bt|D_k}\left(\bt\right)$. We also discuss in Appendix B how to generalize Eq.~\eqref{eq:Bayes} if 
other resources are available, and how the expected values used for the computation of Eq.~\eqref{eq:Bayes} are approximated numerically.

For a given estimation problem and prior $p(\bt)$, the Bayes risk $r\left[h | p(\bt)\right]$ represents our figure of merit (smaller values are better) for experiment-design heuristics.
\section{Experiment design as a RL problem}
Our idea is to train a NN to become an experiment-design heuristic. 
 To this end, we simulate the experiment and the Bayesian update offline
 many times to generate training data [see Fig.~\ref{fig:schemes} (c)] and train the NN with it.  Instead of simulating the experiment, measurement data from an actual experiment could be used to train the NN; note, however, that the calculation of the Bayesian update still depends on the model for the experiment.  Either way, 
once the NN is trained, it represents an adaptive heuristic that can be used as a part of a real Bayesian quantum sensor.

Instead of imitating the behavior of existing heuristics with a neural network, we are interested in creating heuristics which surpass conventional heuristics. To this end, we phrase the problem of experiment-design heuristics in the language of RL which provides us with a suitable framework to optimize heuristics.

RL is an iterative method in which an RL agent, in our case the neural network, learns from training data. These training data are generated in each iteration when the RL agent interacts with the RL
 	environment, as depicted in Fig.~\ref{fig:schemes}(c): The agent chooses an action, i.e., an experiment design, and passes it to the RL environment. In the RL environment, an experiment is simulated according to the action taken by the agent and the measurement outcome is used for the Bayesian update. In return, the agent obtains an observation and a reward. By repeating this cycle several times, training data consisting of actions, observations, and rewards are generated. The learning phase is prescribed by a RL algorithm of choice and generally consists of a combination of learning from experience (using the training data) and random exploration (for more details see next section about algorithms).

The RL framework is based on the agent-environment model and can be understood in the context of Bayesian quantum estimation, cf.~Fig.~\ref{fig:schemes}(c): One \textit{episode} of training data is generated from one simulated Bayesian estimation which consists of a sequence of $k$ simulated experiments including Bayesian updates and experiment designs chosen by the RL agent. The number of experiments $k$ depends on the available resources. In the simplest case, the number of experiments is the limiting resource (referred to as \textit{experiment-limited} in the following), i.e., each episode consists of the same number of experiments, and the Bayes risk is given by Eq.~\eqref{eq:Bayes}. More generally, the available resources set more complicated constraints such that $k$ can vary between different episodes (see Appendix B for a generalization of the Bayes risk for such cases). If a resource is exhausted, the episode ends, the RL environment [see Fig.~\ref{fig:schemes}(c)] is reset to default values (the current posterior is reset to the prior $p(\bt)$ and a new true parameter $\bt$ is sampled from the prior), and another episode starts. Training data for one iteration typically consist of many episodes, e.g., $\sim 10^3$ in our model study below.

Crucial for the success of RL are the observations and rewards for the agent, see Fig.~\ref{fig:schemes}(c). The observations may contain information about the current knowledge 
 $\p{\bt}{D_k}$, about the past, such as prior actions, and about resources, such as the remaining time. 
The reward should reflect the goodness of the behavior (actions) of the agent (larger reward is better) and is used by the RL algorithm to enforce behavior which leads to larger rewards; the RL agent learns from its experience. The negative Bayes risk seems to be an obvious choice for a reward function. However, the computation of the Bayes risk is too time-consuming. Instead, we define the reward after the $k$th experiment as the difference in the traced covariance over the posterior after the $k$th and the $(k-1)$th experiment,
\begin{align}
R(D_k )=\tr\left[\operatorname{Cov}_{\bt|D_{k-1}}\left(\bt\right)\right]-\tr\left[\operatorname{Cov}_{\bt|D_k}\left(\bt\right)\right].\label{eq:reward}
\end{align}
The idea behind this reward function is that (i) Eq.~\eqref{eq:reward} is straightforward to compute in the Bayesian SMC framework, (ii) it reflects the difference in our uncertainty about $\bt$ before and after the current step, and (iii) the expected value of the rewards accumulated from the beginning of an episode with respect to possible measurement outcomes $D_k$ yields the negative Bayes risk (up to a constant), see also Appendix B. For example, in the experiment-limited case with $N$ experiments per episode we have
\begin{align}
\operatorname{E}_{D_N}\left[\sum_{j=1}^N R(D_j)\right] =\text{const}-r\left[h| p(\bt)\right], \label{eq:reward_acc}
\end{align}
where the constant is given by the uncertainty in the prior, $\text{const}=\tr\left[\operatorname{Cov}_{\bt}\left(\bt\right)\right]$. RL has the goal to maximize the expected discounted reward which equals the left-hand side of Eq.~\eqref{eq:reward_acc} because we set the discount factor to one. From Eq.~\eqref{eq:reward_acc} we thus see that RL indeed attempts to minimize the Bayes risk $r\left[h| p(\bt)\right]$.
\section{Algorithms}
We have seen that the problem of finding experiment-design heuristics can be formulated in the language of RL. There is a wide choice of algorithms which can be used to train the agent in a RL setting, ranging from easy-to-implement benchmarks to complicated state-of-the-art reinforcement-learning algorithms. We choose two algorithms from both ends of this spectrum: The cross-entropy method (CEM) for continuous action spaces is a simple evolutionary strategy \cite{knott2016a} which is a good benchmark for more complicated reinforcement learning algorithms \cite{rubinstein1999cross, schulman2015trust,pourchot2018cem}. The algorithm is explained in Fig.~\ref{fig:schemes}(c), for details see Appendix D. Despite its simplicity, we will see that it is capable of yielding very good results.
	
On the other hand, we use trust region policy optimization (TRPO) \cite{schulman2015trust} as implemented in the Python package Stable Baselines \cite{stable-baselines}. We tried several RL algorithms from Stable Baselines and chose TRPO because it showed best stability and performance. However, this was by no means a systematic comparison which would require extensive simulations and hyperparameter optimization for each algorithm. Instead, we tune the hyperparameters of TRPO based on general reasoning \cite{schulman2017deep}. TRPO is an approximation of an iterative procedure for optimizing policies with guaranteed monotonic improvement and  is to date one of the most successful reinforcement learning algorithms  \cite{schulman2015trust}.

The training of the NNs consists of two steps, see Fig.~\ref{fig:schemes} (b):
We initialize the NN using imitation learning (pretraining) as implemented in \cite{stable-baselines}. The idea of imitation learning is to take advantage of an already existing (e.g., manually found) heuristic. The NN is trained to imitate the behavior from episodes created with the existing heuristic, cf.~Fig.~\ref{fig:schemes}(c).  This pretraining step is not strictly necessary but speeds up the training and makes RL more stable. 

However, there might not always be a good heuristic available to imitate. For such cases and for the sake of comparison, we consider CEM. CEM is used to train a neural network from scratch starting from a randomly initialized neural network. 
Once the NN is pretrained, we use RL as a second step.

\begin{figure*}
	\centering
	\includegraphics[width=\linewidth]{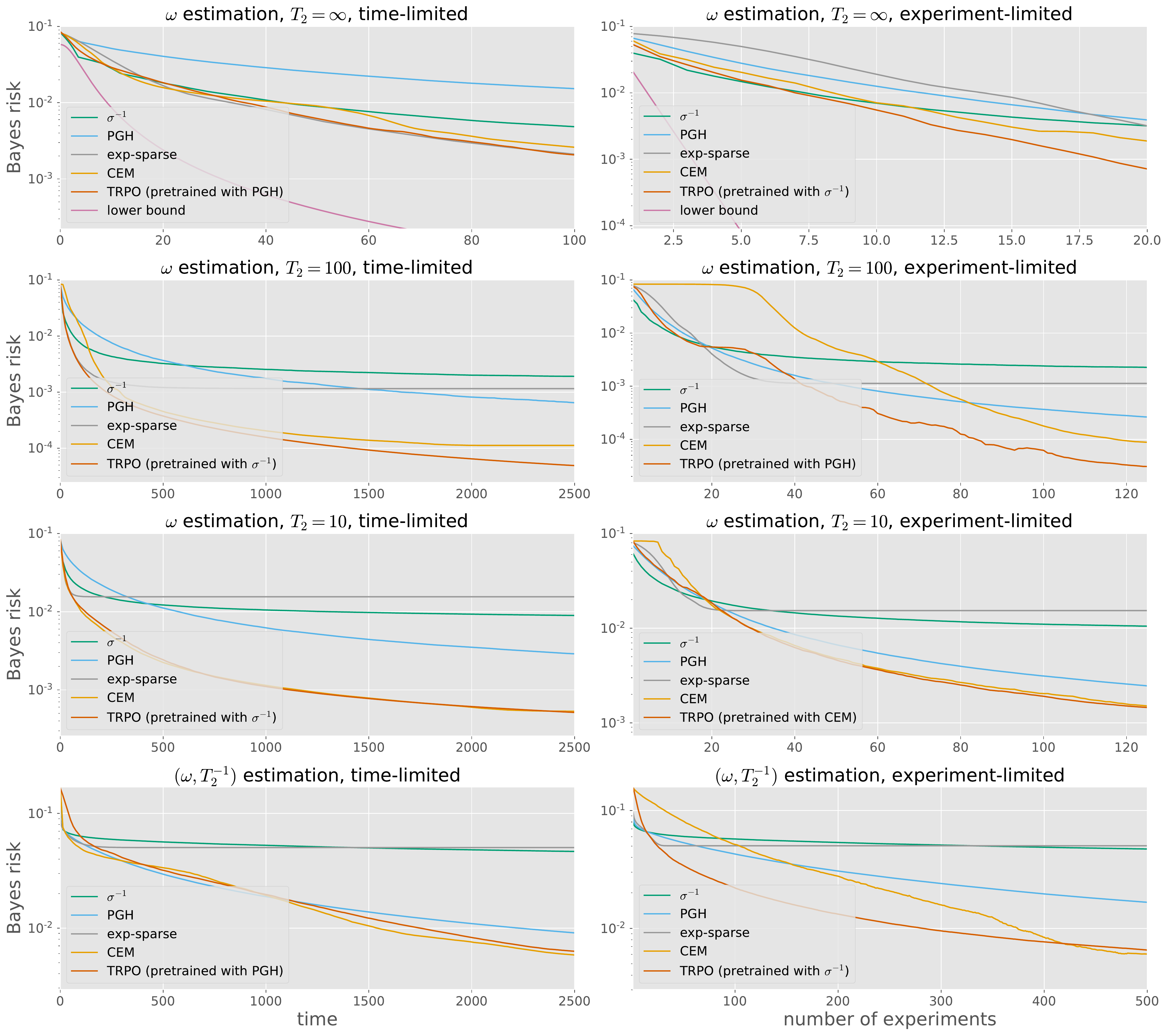}
	\caption{Comparison of the Bayes risk for different experiment-design heuristics. The right column shows Bayesian estimation with a limited number of experiments, the left column with a limit on the available time (limited to the maximal value plotted on the $x$ axis, respectively). We study frequency estimation without (top row) and with $T_2$ relaxation (2nd and 3rd rows, with different values of $T_2$ as stated in the plot titles) as well as the simultaneous estimation of the frequency $\omega$ and relaxation rate $T_2^{-1}$ (bottom row). The Bayes risk is calculated numerically from $10^4$ episodes, see Appendix B for details. TRPO has been pretrained with the heuristic which is specified in brackets in the legends. The lines are linear interpolants to guide the eye.}\label{fig:heuristics}
\end{figure*}
\section{Model study}
We demonstrate our method of creating NN heuristics with an example 
of high practical relevance for magnetic field estimation with nitrogen-vacancy centers with applications in 
single-spin 
magnetic resonance \cite{schmitt2017submillihertz, wang2017experimental, paesani2017experimental, santagati2019magnetic}.
Let us consider a qubit which evolves under the Hamiltonian $H(\omega)=\frac{\omega}{2}\sigma_z$, and we want to estimate the frequency $\omega$. The qubit is prepared in $\ket{+}=\left(\ket{0}+\ket{1}\right)/\sqrt{2}$,
evolves under $H(\omega)$ for a controllable time $t$, and is measured in the $\sigma_x$ basis (assuming a strong projective measurement with outcomes labeled $0$ and $1$ corresponding to a measurement of $\ket{+}$ and $\ket{-}$). Let us further assume that the qubit suffers from an exponential decay of phase coherence, with characteristic time $T_2$. According to the Born rule, the likelihood of finding an outcome $d\in\{0,1\}$ with the $\sigma_x$ measurement can be expressed as \cite{granade2012robust, ferrie2012how},
\begin{align}
p(0|\omega, T_2, t)=\e{-\frac{t}{T_2}}\cos^2\left(\frac{\omega}{2}t\right)+\frac{1-\e{-\frac{t}{T_2}}}{2},\label{eq:likelihood}
\end{align}
for measuring $d=0$,
and $p(1|\omega,t, T_2)=1-p(0|\omega,t, T_2)$ for measuring $d=1$. Eq.~\eqref{eq:likelihood} defines all relevant properties of the experiment. A single experiment design consists of specifying the evolution time $t$. We consider the following estimation problems: (i) the estimation of $\omega$ without decoherence ($T_2=\infty$, see the top row of panels in Fig.~\ref{fig:heuristics}), (ii) the estimation of $\omega$ with known $T_2$ relaxation (we consider this problem twice with different values for $T_2$, see the second and third row of panels in Fig.~\ref{fig:heuristics}), and (iii) the simultaneous estimation of $\omega$ and $T_2^{-1}$, i.e., $\bt=(\omega, T_2^{-1})$ (see the bottom row of panels in Fig.~\ref{fig:heuristics}). In all cases we consider $\omega\in(0,1)$ (making the problem dimensionless). We chose these problems in order to cover wide range of estimation problems; parameters are not cherry-picked and similar results can be expected for different parameter choices (in Section Discussion and Conclusion we discuss limitations of our method).

\begin{figure*}
	\centering
	\includegraphics[width=\linewidth]{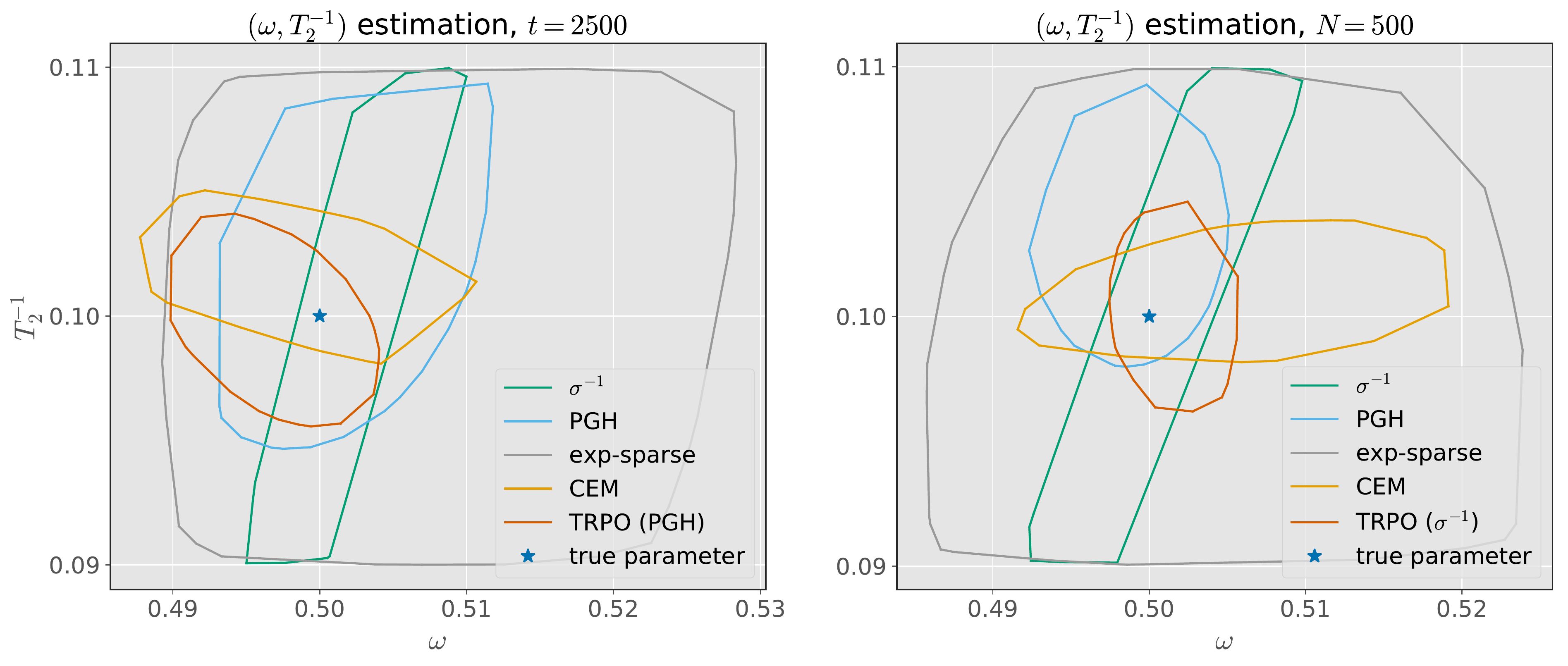}
	\caption{$95\%$ credible regions for multiparameter estimation of $(\omega, T_2^{-1})$. The  credible regions correspond to the uncertainty in the posterior distributions obtained by running one Bayesian estimation (episode) for each heuristic while keeping the true parameter fixed.  All heuristics use a uniform prior for $\omega\in (0,1)$, $T_2^{-1}\in (0.09,0.11)$.}\label{fig:regions}
\end{figure*}

Each estimation problem defines a RL environment [see Fig.~\ref{fig:schemes}(c)] which is either \textit{time-limited} or \textit{experiment-limited}. In the former case, the available time $T$ per episode is limited, while in the latter case the number of experiments $N$ per episode is fixed. The first case is relevant if time is the limiting resource while the second case is relevant if measurements are expensive, for instance, if experiments involve probing sensitive substances such as biological tissue.
In practice, there may be constraints on both, $T$ and $N$, which could be easily taken into account by creating a RL environment accordingly.

As an observation for the NN after the $k$th experiment we choose the expected value $\operatorname{E}_{\bt|D_k }\left(\bt\right)$ and the covariance $\operatorname{Cov}_{\bt|D_k }\left(\bt\right)$ over the posterior (that generalizes
the variance in case of single-parameter estimation), the previous actions from the current episode (maximal 30 actions), and the spent time or the number of experiments in the current episode (for the time-limited or experiment-limited case, respectively). 

Several heuristics have been developed for estimation problems (i) and (ii). As an example for a non-adaptive strategy, we consider a heuristic which chooses exponentially sparse times \cite{ferrie2012how}, $t_k=\left(9/8\right)^k$, denoted as \textit{exp-sparse} heuristic in the following.

Further, we consider two adaptive heuristics: We define the first one as $t_k=\tr\left[\text{Cov}_{\bt|D_{k-1}}\left(\bt\right)\right]^{-1/2}$ and we will call it \emph{$\sigma^{-1}$} heuristic. This represents a generalization to multiparameter estimation of a heuristic which was derived for estimation problem (i) ($\omega$ estimation, $T_2\rightarrow\infty$)  by Ferrie \textit{et al.}~\cite{ferrie2012how}. For single-parameter estimation, the $\sigma^{-1}$ heuristic chooses the times $t_k$ as the inverse standard deviation of $\bt$ over the posterior and is optimal in the greedy sense and only in the asymptotic limit $N\rightarrow\infty$.

The second adaptive heuristic that we consider is the \textit{particle guess heuristic} (PGH) \cite{wiebe2014hamiltonian}. It is based on the SMC framework which uses a particle filter to represent probability distributions such as $p(\bt|D_k)$ \cite{huszar2012adaptive} (see Appendix A). PGH chooses times as the inverse distance of two particles $\bt_1, \bt_2 \in \Theta$ sampled from $p(\bt|D_{k-1})$, $t_k=\lVert\bt_1-\bt_2\rVert_2^{-1}$. In case of single-parameter estimation, PGH is a proxy for the $\sigma^{-1}$ heuristic but it is faster to compute (given the particle filter) and introduces additional randomness (compared with the $\sigma^{-1}$ heuristic).

Let us turn to the results depicted in Fig.~\ref{fig:heuristics}.
In all examples, we consider uniform priors for $\omega\in(0,1)$ and, in case of multiparameter estimation, also for $T_2^{-1}\in(0.09,0.11)$. Uniform priors are the logical choice when no prior knowledge of the parameters is available. Also note that over the course of a Bayesian estimation many different posteriors (i.e., priors for future experiments) are reached. For the multiparameter estimation examples, we consider, instead of one experiment, $100$ independent and identical experiments in each step in order to facilitate the estimation of $T_2^{-1}$, and we also give the averaged outcome of these experiments as an additional observation to the RL agent.

We
considered
three different heuristics for pretraining TRPO: the $\sigma^{-1}$ heuristic, PGH, and a CEM-trained NN,
but plot only the
results 
for the best of these three.
For the data in Fig.~\ref{fig:heuristics},  CEM is implemented with one hidden layer with 16 neurons, TRPO with a NN with two hidden layers with 64 neurons each.
The NN heuristics outperform the conventional heuristics in all examples. Note that TRPO performs better than CEM when the limiting resource is used only partly. In the presence of $T_2$ relaxation, times which exceed $T_2$ tend to yield no information, which explains why the Bayes risk saturates for the exp-sparse heuristic. The largest advantage of a NN heuristic is found for the example of time-limited $\omega$ estimation with $T_2=10$. Compared with PGH, we find an improvement in the Bayes risk by more than one
order of
magnitude.

Generally, the performance of NN heuristics is remarkable given that (for the single-parameter estimation problems considered in Fig.~\ref{fig:heuristics}) the conventional heuristics such as PGH are used in experiments \cite{wang2017experimental, paesani2017experimental, lumino2018experimental, santagati2019magnetic} and are considered to be the best practical choice with near-optimal performance \cite{wiebe2014hamiltonian, wiebe2016efficient}. This raises questions about the ultimate performance which could be achieved with an optimal heuristic. However, known lower bounds for the Bayes risk such as the Bayesian Cram\'er--Rao bound \cite{vantrees1968detection, gill1995applications} usually become very loose when they are optimized with respect to heuristics (for a detailed discussion see Appendix C). In Fig.~\ref{fig:heuristics}, we provide lower bounds for estimation problem (i). Remarkably, the lower bound in the experiment limited case can in principle be saturated using a phase estimation protocol based on the semiclassical quantum Fourier transform \cite{childs2000quantum}. However, this protocol requires the implementation of $\sigma_z$-rotations dependent on previous measurement outcomes. This is not possible in our example because $\sigma_z$-rotations are not part of the available resources. Lower bounds provide an ultimate limit to the performance but, considering their their looseness, leave us in the dark about the true potential of experiment-design heuristics.

Another question concerns how experiment designs, i.e., the evolution times, chosen by the NN heuristics compare with those from the conventional heuristics. In Appendix E, we compare the distribution of experiment designs for the adaptive heuristics used in Fig.~\ref{fig:heuristics}.

While the Bayes risk is used to compare the expected (average) performance of heuristics, it is an important advantage of the Bayesian approach over the frequentist one that it provides credible regions as a practical tool for comparing single runs of parameter estimation (episodes) \cite{qinfer, granade2017qinfer}. Let us revisit the multiparameter problem discussed in the two bottom panels of Fig.~\ref{fig:heuristics}. This time, we run only one Bayesian estimation of $(\omega, T_2^{-1})$ with each heuristic and visualize their performance by plotting $95\%$ credible regions \cite{qinfer, granade2017qinfer}, see Fig.~\ref{fig:regions}.  Note that since the Bayesian estimation is subject to fluctuations (such as stochastic measurement outcomes), the shape of credible regions fluctuates between different estimations. For the specific example shown, we see that TRPO provides the smallest uncertainties compared with the
other heuristics as judged by the area of the credible regions.

\section{Discussion and conclusion}
We propose to use neural networks (NNs) as experiment-design heuristics and to train, i.e.~optimize, them using reinforcement learning (RL).  The properties of the estimation problem and the quantum experiments, and  the availability of resources are taken into account during the training which results in tailored NN-based experiment-design heuristics.
	
A big advantage of our method lies in its versatility and  adaptivity as we will discuss in the following. On the computational side, limitations are mostly related to slow numerical Bayesian updates or a large number of calls to the heuristic which slows down the generation of training data and thus increases the run time for training the neural networks. In cases when Bayesian estimation involves many experiments, one can partly sacrifice adaptivity and switch to calling the NN heuristic only every $n$th experiment in which case the NN would design the next $n$ experiments rather than just one experiment and, thus, the number of calls to the heuristic is in principle under our control. On the other hand, efficient numerical Bayesian updates are a prerequisite for efficient Bayesian quantum estimation in the first place. Therefore, as a rule of thumb, our method can be applied to adaptive Bayesian quantum estimation whenever the Bayesian updates can be computed efficiently. This applies to all kinds of Bayesian quantum sensors and arbitrary constraints on the available resources.
	
When it comes to the performance of NN heuristics we obviously cannot guarantee that RL always succeeds or that in practice the NN heuristics will be more successful than existing heuristics. However, modern RL algorithms such as trust region policy optimization are known to perform well for a wide range of problems \cite{schulman2015trust}. While we demonstrate the success of TRPO for a total of eight estimation problems in the context to phase estimation including examples of multiparameter estimation, we can be optimistic that RL will also work for other estimation problems. Also note that despite phase estimation being one of the best studied examples of Bayesian quantum estimation \cite{ferrie2012how, granade2012robust, wiebe2014hamiltonian, paesani2017experimental, lumino2018experimental, santagati2019magnetic}, our NN heuristics are able to outperform the established heuristics.	
 
The practical success of adaptive Bayesian estimation often depends on the run time of data processing (Bayesian update, choice of an adaptive experiment design). 
Using NNs as experiment-design heuristics introduces a computation
overhead compared with the fastest existing experiment-design heuristics
such as PGH. On the other hand, improvements in measurement
precision achieved by the NN heuristics may easily outweigh the
drawback of an increased run time. This trade-off has to be
assessed dependent on the estimation problem and its concrete
realization. In our model study, the run time for one call to a NN
heuristic is always shorter than the corresponding numerical Bayesian
update (single-core computation in both cases). For more complicated
modeling of the sensor or for an estimation up to a larger precision,
the run time of the Bayesian update increases 
even more
and we expect that the run time will be dominated by the numerical
Bayesian update. Moreover, the effective run time of the NN per experiment
could be reduced by using smaller NNs or, as discussed before, calling the NN heuristics only every $n$th experiment. Alternatively, it could be interesting to use NN heuristics as ``warm-up heuristics'', i.e., using another heuristic afterwards,   especially if faster heuristics are available which are known to be asymptotically optimal. Further, it should be noted that run-time considerations can be less important when other resources are scarce, for example, when experiments involve probing sensitive substances we may want to minimize the number of experiments rather than time (which is the reason why we consider not only time-limited but also experiment-limited estimation problems).

In conclusion, we propose and demonstrate
a machine learning method to create fast and strong experiment-design heuristics for Bayesian quantum estimation. The method uses imitation and reinforcement learning for training NNs to become experiment-design heuristics. In order to make the method independent of the availability of an expert heuristic for imitation learning, we show that expert heuristics can be found with an evolutionary strategy (the cross-entropy method).

We provide the complete source code \footnote{Source code and data are available at \href{https://doi.org/10.6084/m9.figshare.11927043}{https://doi.org/10.6084/m9.figshare.11927043}.} used for this work in order to facilitate the application of the presented method in experiments and to related problems such as the detection of time-dependent signals \cite{isard1998condensation, granade2017qinfer} and adaptive Bayesian state tomography \cite{granade2016practical, granade2017practical}.

\begin{acknowledgments}
	L.F.~thanks Paul Knott for helpful discussions about RL.
	L.F.~and D.B.~acknowledge
	support from the Deutsche Forschungsgemeinschaft (DFG), Grant No.~BR
	\mbox{5221/1-1}. L.F.~acknowledges financial support from the European Research Council (ERC) under the Starting Grant \mbox{GQCOP} (Grant No. 637352), and from the Austrian Science Fund (FWF) through SFB BeyondC (Grant No.~F7102). J.S.~was supported by the Otto A.~Wipprecht Stiftung. 
\end{acknowledgments}

\setcounter{equation}{0}
\renewcommand\theequation{A.\arabic{equation}}
\section{Appendix A: Bayes' law and the sequential Monte Carlo algorithm}
Bayes' law for updating our knowledge of $\bt\in\Theta$ according to the measurement outcome $d_k$ of the $k$th experiment is given by
\begin{align}
p(\bt|D_k)=\frac{p(d_k|\bt)p(\bt|D_{k-1})}{p(d_k)}, \label{eq:Bayes_law}
\end{align}
where $p(\bt|D_k)$ is our updated knowledge (posterior), $p(\bt|D_{k-1})$ is our knowledge prior to the $k$th experiment, $p(d_k|\bt)$ is the likelihood of measuring $d_k$, and $p(d_k)$ is a normalization, $p(d_k)=\operatorname{E}_{\bt}\left[p(d_k|\bt)\right]$.
The exact solution for the Bayesian update is generally intractable. Instead, we use an inference algorithm based on the sequential Monte Carlo algorithm \cite{doucet2009tutorial, huszar2012adaptive, granade2012robust}. The idea is to represent the probability distributions $p(\bt)$ and $p(\bt|D_k)$ by a discrete approximation $\sum_{j=1}^{n}w_j\delta(\bt-\bt_j)$ with $n$ so-called particles with positive weights $w_j$ and positions $\bt_j\in\Theta$. Then, for a Bayesian update, we only need to update the weights of each particle by calculating $p(d_k|\bt_j)$. This means 
that 
for the $j$th particle we have to simulate the experiment for $\bt=\bt_j$ and use the Born rule to find $p(d_k|\bt_j)$. The expected value in the definition of $p(d_k)$ reduces to a simple sum over the particles of the prior.

The particle locations need to be  resampled if too many weights are close to zero, i.e., the particle filter of $p(\bt|D_k)$ is impoverished. We use Qinfer's \cite{qinfer} implementation of the Liu--West resampling algorithm \cite{liu2001combined} (with default parameter $a=0.98$ \cite{qinfer})
with $n=2\times 10^3$ particles for RL environments without decoherence and for RL environments with multiparameter estimation. For the environments with $\omega$ estimation and finite $T_2$ we use  $n=2\times 10^4$ particles. Particle numbers are chosen sufficiently large in order to provide an accurate approximation of $p(\bt|D_k)$ (i.e., when increasing the number of particles, the Bayes risk does not change). Generally, there is a tradeoff between a smaller number of particles, which leads to reduced run times but sacrifices accuracy, and larger number of particles, which provides a more accurate approximation at the cost of an increased run time.
\setcounter{equation}{0}
\renewcommand\theequation{B.\arabic{equation}}
\section{Appendix B: Bayes risk}
Let us consider a sequence of $k$ experiments designed with the experiment-design heuristic $h$. Let $\hat{\bt}_k: D_k\mapsto \hat{\bt}_k(D_k)$ be an estimator of $\bt$ after $k$ experiments, and let $L\left[\hat{\bt}_k(D_k)| \bt\right]$ be a loss function which quantifies the deviation of $\hat{\bt}_k(D_k)$ from $\bt$. For an experiment-design heuristic $h$, the risk $s$ of $\hat{\bt}_k$ is defined as (usually denoted by $R$ which denotes the reward in this work)
\begin{align}
s(\hat{\bt}_k, h |\bt)=\operatorname{E}_{D_k|\bt}\left(L\left[\hat{\bt}_k(D_k)| \bt\right]\right). \label{eq:risk}
\end{align}
Note that the experiment-design heuristic $h$ determines the experiment designs $E_k$ which influence the estimate $\hat{\bt}_k(D_k, E_k)$. However, in order to simplify notation we write $\hat{\bt}_k(D_k)$ instead of $\hat{\bt}_k(D_k, E_k)$.
For a  definition of the expected value $\operatorname{E}_{D_k|\bt}$, see in the main text after Eq.~\eqref{eq:Bayes}. Eq.~\eqref{eq:risk} represents the risk typically associated with a choice of an estimator $\hat{\bt}_k$ and corresponds to the expected (with respect to measurement outcomes $D_k$) loss. The Bayes risk represents a way to additionally take into account prior knowledge of $\bt$ in form of a probability distribution $p(\bt)$ (the prior) on $\Theta$. Given a prior $p(\bt)$, the Bayes risk is defined as
\begin{align}
r\left[\hat{\bt}_k, h| p(\bt)\right] =  \operatorname{E}_\bt\left[s(\hat{\bt}_k, h |\bt)\right]. \label{eq:Bayes_usual}
\end{align}
A common choice for a loss function is the quadratic loss $L\left[\hat{\bt}_k(D_k)| \bt\right]=\pnorm{\hat{\bt}_k(D_k)-\bt}{2}^2$  \cite{granade2012robust}. Then, the estimator which minimizes the Bayes risk, the Bayes estimator, is given by the expectation over the posterior, $\hat{\bt}_k (D_k)=\operatorname{E}_{\bt|D_k}(\bt)$. For the quadratic loss, the risk is also known as mean squared error, and the Bayes estimator is also known as minimum mean square error estimator.
\subsection{Derivation of Eq.~(1) in the main text}
 We find from  Eq.~\eqref{eq:Bayes_usual},
\begin{align}
r\left[\hat{\bt}_k,h| p(\bt)\right]&=\operatorname{E}_\bt\left[\operatorname{E}_{D_k|\bt}\left(\pnorm{\hat{\bt}_k(D_k)-\bt}{2}^2\right)\right]\label{eq:1}\\
&=\operatorname{E}_{D_k}\left[\operatorname{E}_{\bt|D_k}\left(\pnorm{\hat{\bt}_k(D_k)-\bt}{2}^2\right)\right],\label{eq:2}
\end{align}
where we used the definition of the expected values together with Bayes' law $p(D_k|\bt)=\frac{p(\bt|D_k)p(D_k)}{p(\bt)}$.
Next, we insert the Bayes estimator in Eq.~\eqref{eq:2} in order to write the Bayes risk solely as a function of the heuristic $h$ which corresponds to Eq.~\eqref{eq:Bayes} in the main text:
\begin{align}
r\left[h| p(\bt)\right]&=\operatorname{E}_{D_k}\left[\operatorname{E}_{\bt|D_k}\left(\pnorm{\operatorname{E}_{\bt|D_k}(\bt)-\bt}{2}^2\right)\right]\label{eq:3}\\
&=\operatorname{E}_{D_k}\left(\sum_j\operatorname{Var}_{\theta_j|D_k}\left[\theta_j\right]\right) \label{eq:4}\\
&=\operatorname{E}_{D_k}\left[\tr\left(\operatorname{Cov}_{\bt|D_k}\left[\bt\right]\right)\right].\label{eq:5}
\end{align}
\subsection{The Bayes risk in the presence of limited resources}
As discussed in the main text, the available resources for experiment designs may lead to episodes with different numbers of experiments. A Bayes risk which corresponds to the situation that the available resources are exhausted can easily be generalized from Eq.\eqref{eq:Bayes} in the main text as
\begin{align}
r\left[h, p(\bt)\right] = \operatorname{E}_{D_\text{end}}\left(\tr\left[\operatorname{Cov}_{\bt|D_\text{end}}\left(\bt\right)\right]\right), \label{eq:Bayes_redefined}
\end{align}
where $D_\text{end}$ denotes all measurement outcomes for an episode. This means, the expectation $\operatorname{E}_{D_\text{end}}$ must be taken with respect to all data sets $D_k$ which are compatible with the available resources.
\subsection{Numerical Approximation of the Bayes risk}
Let us first consider how to approximate Eq.~\eqref{eq:3} which is relevant for the experiment-limited case considered in our model study. 
The Bayes risk in Eq.~\eqref{eq:3} involves expected values over $p\left(\bt|D_k \right)$ and $p\left(D_k\right)$. The posterior $p\left(\bt|D_k \right)$ is represented in the SMC framework by a particle filter with weights $w_j$ and particle locations $\bt_j\in \Theta$ which allows us to approximate the expected value of a function $f(\bt)$ as $\operatorname{E}_{\theta|D_k }\left[f(\bt)\right]=\sum_j w_j f(\bt_j)$. The same approximation is used to compute the reward. Note that we set 
\begin{align}
\tr\left(\operatorname{Cov}_{\bt}\left[\bt\right]\right)=\text{const} \label{eq:const}
\end{align}
to a constant numerical value in order to avoid fluctuations of the reward originating from numerical uncertainties in the particle filter representing the prior $p(\theta)$.

The expected value over $p\left(D_k\right)$ is approximated as a sample mean: We sample  $10^4$ episodes with the RL environment. To each episode corresponds a series of measurement outcomes $D_k$ and we can calculate $\tr\left[\operatorname{Cov}_{\bt|D_k}\left(\bt\right)\right]$. Note that the properties of the RL environment ensure that an episode with data $D_k$ is sampled with probability $p(D_k)$. Then, we can approximate the expected value over $p(D_k)$ as the mean
$\frac{1}{10^4}\sum_{D_k}\tr\left[\operatorname{Cov}_{\bt|D_k}\left(\bt\right)\right]$, where the sum runs over all $10^4$ sampled series of measurement outcomes $D_k$.
 
The time-limited case we consider in our model study is an example where episodes can have different numbers of experiments. Moreover, we defined the time-limited case such that an episode ends when the agent has consumed more time than the given time limit. This means with the last experiment the RL agent (as well as the other heuristics) can go beyond the time limit. For the comparison of heuristics, this is not relevant mainly because all other experiment designs must lie within the time limit.

\begin{figure}
	\centering
	\includegraphics[width=\linewidth]{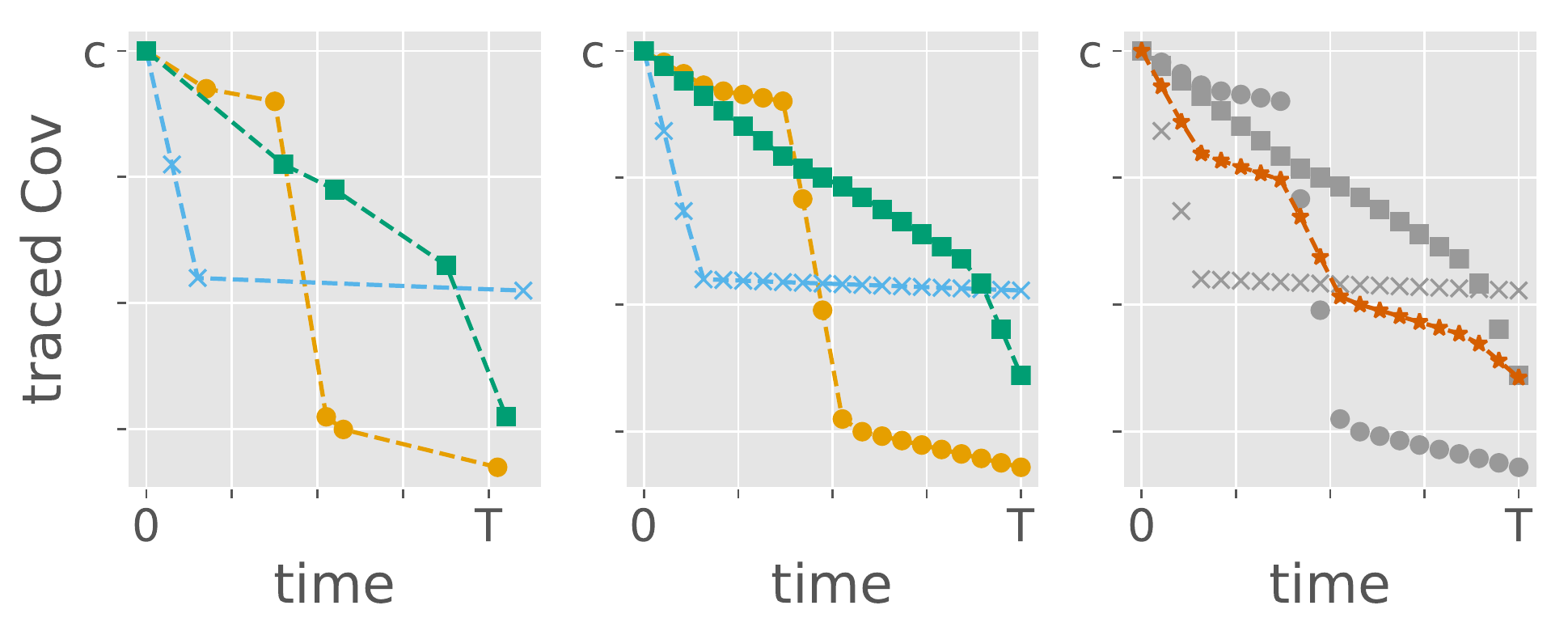}
	\hspace*{0.5cm} (a) \qquad  \qquad \quad \qquad (b) \qquad \quad \qquad \qquad (c)
	\caption{Illustration of Bayes risk calculation for time-limited environments.  Panel (a) shows 3 time series each of which corresponds to one episode (i.e., one Bayesian estimation): the ``$0$th'' data point at time $0$ corresponds to the uncertainty $c$ in the prior, see Eq.~\eqref{eq:const}, and the $k$th data point in each time series corresponds to the $k$th experiment of that episode; its $x$ coordinate is the accumulated time used for the first $k$ experiments, and its $y$ coordinate is the traced covariance $\tr\left[\operatorname{Cov}_{\bt|D_k}\left(\bt\right)\right]$. The time limit is $T$ which is overstepped only with the last experiment of each episode. Linear interpolation of each time series (dashed lines) and evaluation of the linear interpolants at equally spaced times spanning $[0,T]$ yields panel (b), depicted with $20$ equally spaced times. Then, we can take a time-wise mean of the data, i.e., for each of the $20$ equally spaces times, we calculate the mean over all time series. This gives us the mean of the interpolated traced covariance [depicted as red stars in Panel (c)] which serves as an approximate time-resolved Bayes risk.}\label{fig:explain}
\end{figure}

For the time-limited case, it would be insightful to have a time-resolved Bayes risk in analogy to the experiment-limited case, where for every number of experiments, we can compute a Bayes risk. Fig.~\ref{fig:explain} illustrates how we approximate such a time-resolved Bayes risk (actually, we average over $10^4$ episodes and use $200$ equally spaced times for interpolation). This approximation of the time-resolved Bayes risk is used for the time-limited cases shown in Fig.~\ref{fig:heuristics} in the main text.
\section{Appendix C: Lower bounds for the Bayes risk}
Since we are interested in optimizing experiment-design heuristics, we would like to know whether our heuristics are actually close to optimal or if we miss out on potentially much larger improvements. To this end, a lower bound $L$ for the Bayes risk would be useful such that $r\left[h| p(\bt)\right]\geq L$ for all heuristics $h$. A well-known lower bound for the Bayesian risk is the Bayesian Cram\'er--Rao bound (BCRB) (also known as van Trees inequality \cite{vantrees1968detection}), cf. Eq.~(10) in Ref.~\cite{gill1995applications}:
\begin{align}
r\left[h| p(\bt)\right]\geq \frac{1}{\operatorname{E}_\bt\left[\tr\left(I_\bt\left[p(D_k|\bt)\right]+I_\bt\left[p(\bt)\right]\right)\right]},\label{eq:BCRB}
\end{align}
where we defined
\begin{align}
	I_\bt[f(\bt)]=\operatorname{E}_{D_k|\bt}\left(\left[\nabla_\bt \log f(\bt)\right]^\intercal \left[\nabla_\bt \log f(\bt)\right]\right) \label{eq:Fisher}
\end{align}
with $\nabla_\bt$ the vector differential operator with respect to $\bt$. In Eq.~\eqref{eq:BCRB}, $I_\bt\left[p(D_k|\bt)\right]$ is the Fisher information matrix, and $I_\bt\left[p(\bt)\right]$ is a contribution due to the initial prior $p(\bt)$. We can see that the BCRB depends on the experiment-design heuristic since the Fisher information matrix depends on the likelihood and thus on the experiment designs. Therefore, a lower bound $L$ could be defined as
\begin{align}
L&=\inf_h\frac{1}{\operatorname{E}_\bt\left[\tr\left(I_\bt\left[p(D_k|\bt)\right]+I_\bt\left[p(\bt)\right]\right)\right]} \label{eq:lower1}\\
&=\frac{1}{\sup_h\operatorname{E}_\bt\left[\tr\left(I_\bt\left[p(D_k|\bt)\right]\right)\right]+\operatorname{E}_\bt\left[\tr\left(I_\bt\left[p(\bt)\right]\right)\right]}\label{eq:lower2}
\end{align}
where we find the infimum (supremum) with respect to all heuristics $h$, which respect the available resources, such that $r\left[h| p(\bt)\right]\geq L~\forall h$.

\begin{figure}[h!]
 	\centering
 	\includegraphics[width=0.6\linewidth]{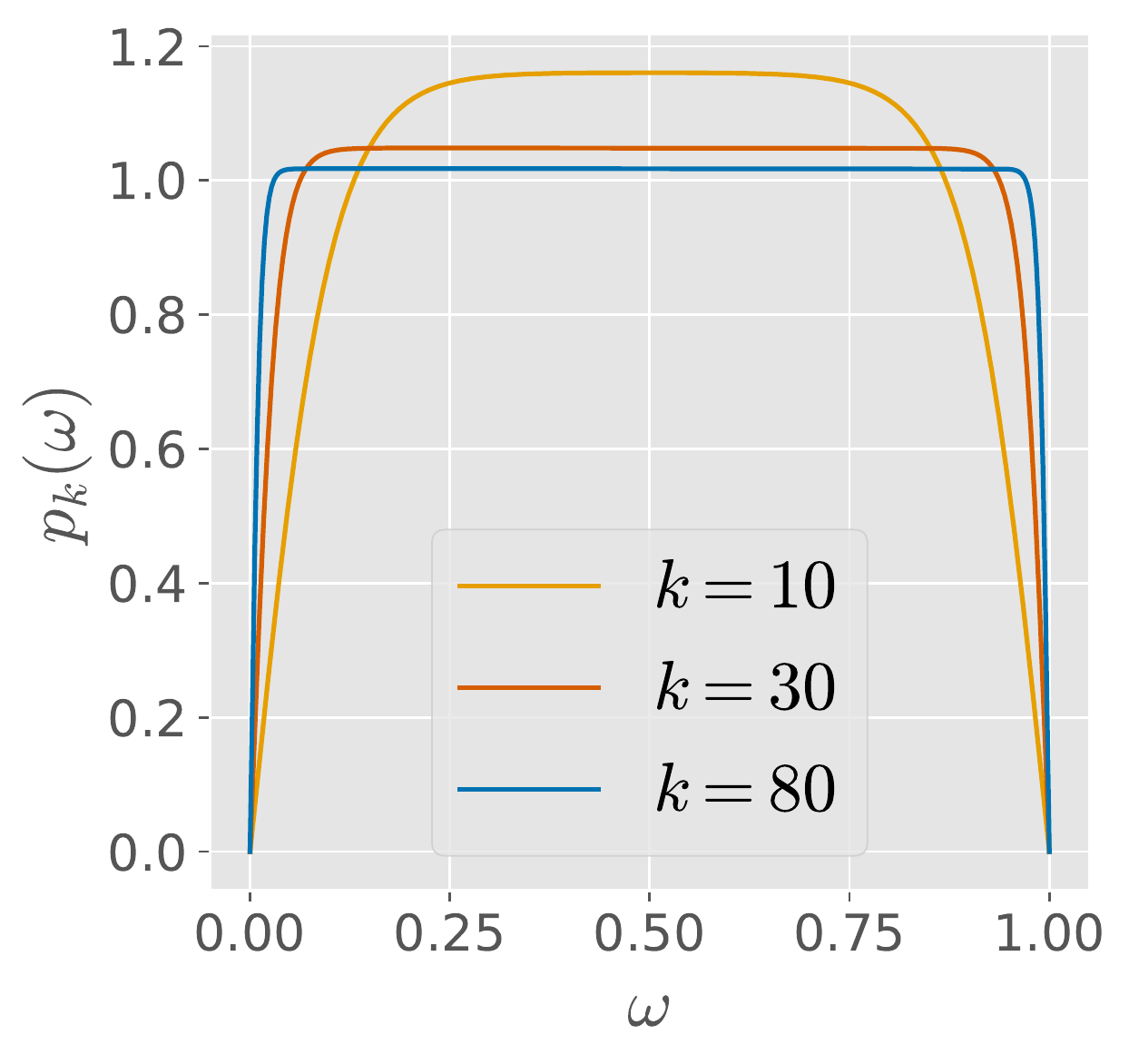}
 	\caption{Analytic approximations of a flat prior which vanishes at the endpoints of its domain. The plotted functions correspond to Eq.~\eqref{eq:priors} with $a=0$ and  $b=1$ for different values of $k$.}\label{fig:priors}
 \end{figure}

Before we discuss the problems related to the lower bound \eqref{eq:lower1}, we have to address another issue related to the BCRB. The BCRB can be applied only if some regularity conditions are fulfilled \cite{gill1995applications}; in particular, the prior $p(\bt)$ must  converge towards zero at the endpoints of domain $\Theta\ni\bt$, where $\Theta$ is assumed to be a finite interval for each parameter $\theta_j$. Therefore, the BCRB cannot be applied directly in our case because we use uniform priors. However, the particle filter used to represent a uniform prior can be seen as an equally good approximation of a prior which vanishes at the endpoints but is otherwise flat. For example, the following ansatz for a prior (up to normalization)
\begin{align}
p_k(\theta)\propto\frac{2}{\left(1 +\e{-2 k (x - a)} \right) \left(1 +\e{2 k (x - b)}\right)}- 1 \label{eq:priors}
\end{align}
converges toward a uniform prior on $(a,b)$ for $k\rightarrow \infty$. Fig.~\ref{fig:priors} shows $p_k(\theta)$ for different values of $k$.

Let us compute the BCRB for estimation problem (i), frequency estimation without dephasing. In case of single-parameter estimation, the Fisher information matrix reduces to a scalar, the Fisher information. Using the likelihood in Eq.~\eqref{eq:likelihood} with $T_2\rightarrow \infty$, we find the Fisher information for the first experiment,
\begin{align}
I_\omega\left[p(D_1|\omega)\right]=t_1^2, \label{eq:fisher_1}
\end{align}
where $t_1$ is the evolution time of the experiment. The Fisher information for $k$ experiments designed with an adaptive heuristic $h$ reads
\begin{align}
I_\omega\left[p(D_k|\omega)\right]=\sum_{j=1}^{k}I_\omega\left[p(D_k^{(j)}|\omega)\right], \label{eq:fisher_total}
\end{align}
where $D_k^{(j)}$ denotes the first $j$ coefficients (measurement outcomes) of $D_k$. We see from Eq.~\eqref{eq:fisher_1} that $I_\omega\left[p(D_k^{(j)}|\omega)\right]=t_{D_k^{(j)}}^2$, where 
$t_{D_k^{(j)}}$ is the evolution time of the $j$th experiment when we observe data $D_k$. This means that the total Fisher information in Eq.~\eqref{eq:fisher_total} does not depend on the parameter $\omega$. Thus, taking the expected value in Eq.~\eqref{eq:lower2} is trivial.
Similarly, the maximization with respect to heuristics $h$ is straightforward: In the time-limited case with time limit $T$, it is easy to see that the sum in Eq.~\eqref{eq:fisher_total} is maximized by choosing a single experiment of time $T$. Then, the supremum with respect to heuristics $h_T$, i.e., heuristics which consume at most time $T$, is given by
\begin{align}
\sup_{h_T}	I_\omega\left[p(D_k|\omega)\right]=T^2,
\end{align}
which would correspond to a heuristic $h_T$ that deterministically chooses a single experiment with evolution time $T$. The lower bound for the Bayes risk reads
\begin{align}
L_T=\frac{1}{T^2+\operatorname{E}_\bt\left[\tr\left(I_\bt\left[p(\bt)\right]\right)\right]} \label{eq:lt}
\end{align}
This also shows that optimizing the BCRB does not necessarily provide useful insights when it comes to finding good experiment designs; in case of qubit measurements, performing only a single experiment yields at best one bit of information about $\bt$.

In the experiment-limited case, where heuristics are allowed to consume arbitrary amounts of time, minimizing the BCRB is pointless because $I_\omega\left[p(D_k|\omega)\right]$ becomes infinite if $t_k\rightarrow\infty$. Alternatively, an information-based lower bound can be derived. Similarly to Refs.~\cite{childs2000quantum, ferrie2012how}, we write $\omega$ in its binary expansion,
\begin{align}
	.\omega_1\omega_2\omega_3\dotsm.
\end{align}
$k$ experiments yield at best $k$ bits of information about $\omega$ such that we know the first $k$ bits $\omega_1, \dotsc, \omega_k$ with certainty. Then, possible $\omega$ values are restricted to an interval $[\omega_\text{min},\omega_\text{max}]$ whose end points are obtained by setting all remaining bits to zero or all to one, i.e., $\omega_\text{min}=.\omega_1 \dotsm \omega_k$ and $\omega_\text{max}=.\omega_1 \dotsm \omega_k111\dotsm $, such that $\omega_\text{max}-\omega_\text{min}=2^{-k}$. Since we use uniform priors in this work, we assume a uniform probability density for $\omega$ on this interval. This means our best estimate is to choose the mean of this interval, $\hat{\omega}_k=.\omega_1\dotsm\omega_k1$.  This yields the following Bayes risk
\begin{align}
	L_k&=\operatorname{E}_\omega\left[\operatorname{E}_{D_k|\omega}\left(\pnorm{\hat{\omega}_k-\omega}{2}^2\right)\right]\\
	&=\operatorname{E}_\omega\left[\left(\hat{\omega}_k-\omega\right)^2\right]\\
	&=2^{-k}\int_{\omega_\text{min}}^{\omega_\text{max}}\text{d} \omega\left(\hat{\omega}_k-\omega\right)^2\\
	&=\frac{2^{-2(k+1)}}{3}.\label{eq:lk}
\end{align}
Remarkably, this Bayes risk is achievable using an adaptive phase estimation protocol which is equivalent to a semiclassical quantum Fourier transform \cite{childs2000quantum}. However, in order to implement the semiclassical quantum Fourier transform, one needs to implement $\sigma_z$-rotations dependent on previous measurement ourcomes (or equivalently one needs a freedom in the choice of the measurement basis) which is not allowed in our scenario and therefore cannot be realized using our available resources.

To sum up, this illustrates how difficult it is to find meaningful lower bounds for the Bayes risk of experiment-design heuristics which are subject to certain available resources. In Fig.~\ref{fig:heuristics} in the main text, we use $L_k$ from \eqref{eq:lk} and $L_T$ from Eq.~\eqref{eq:lt} (with a prior as specified in Eq.~\eqref{eq:priors} with $k=30$)  as lower bounds for the Bayes risk for estimation problem (i).
\section{Appendix D: Details on the training of the neural networks} 
Our results are obtained with NumPy 1.17.3 \cite{numpy}, QInfer 1.0a1 \cite{qinfer}, Gym 0.14.0 \cite{gym}, PyTorch 1.3.1 \cite{paszke2019pytorch}, Stable Baselines 2.9.0 \cite{stable-baselines}, and Tensorflow 1.14.0 \cite{tensorflow} libraries for  Python 3.6.7.
CEM and TRPO use fully connected neural networks (CEM: one hidden layer with 16 neurons, TRPO: two hidden layers with 64 neurons each).
\subsection{The combination of CEM and TRPO}
While CEM takes into account only the accumulated reward of full episodes, RL uses all the training data consisting of actions with corresponding observations and rewards. This allows TRPO to take into account the performance of single actions in order to improve the overall performance (at the end of episodes).  If we pretrain TRPO with a CEM-trained NN (as a known heuristic), we obtain a purely machine learning based method for finding strong NN heuristics. The advantage of this two-step procedure over using only one of the algorithms is that we can use CEM for exploring the policy space (for our RL environments, CEM has proven to be better than TRPO in this respect) and TRPO for optimizing the heuristic further. A further speed-up and possibly an improvement of this method could be achieved by passing the neural network directly from CEM to TRPO (avoiding imitation learning). However, in the current implementation, this is not possible because the neural networks used for CEM and TRPO are of different shape and not compatible.
\subsection{Cross-entropy method} 
The input layer of the NN is defined by the observation. The output layer is determined by the number of actions (one action: time) and we choose 16 neurons in the hidden layer. The layers are fully connected. The hidden layer has the rectified linear unit (ReLU) as its activation function and the output layer has the softmax function as its activation function \cite{nielsen2015neural}. 

A schematic representation of the algorithm is given in Fig.~\ref{fig:schemes}(d) in the main text. The weights of the neural network form a vector $\boldsymbol{x}$. A generation is sampled from a Gaussian distribution $\boldsymbol{x_i}\sim\mathcal{N}(\boldsymbol{\mu},\Sigma)$ with mean $\boldsymbol{\mu}$ and covariance $\Sigma=\one/2 $. Before the first iteration, we sample $\boldsymbol{\mu}$ from $\mathcal{N}(\boldsymbol{0},\Sigma)$. A generation consists of $K=100$ individuals (=NNs). By running one episode (interacting with the RL environment) with each NN, we determine the $K_\epsilon=10$ fittest individuals $(\boldsymbol{x_1}, \dotsc, \boldsymbol{x}_{K_\epsilon})$ as those with the largest reward accumulated over an episode.
This concludes one iteration, the next generation is sampled from the distribution $\mathcal{N}(\boldsymbol{\mu}_\text{new},\Sigma)$ with 	$\boldsymbol{\mu}_\text{new}=\frac{1}{K_\epsilon}\sum_{j=1}^{K_\epsilon}\boldsymbol{x_i}$. We run CEM for $N=1000$ iterations. The final solution is given by $\boldsymbol{\mu}_\text{new}$ calculated in the last iteration.
CEM is used to find  $5$ NNs for each RL environment and we plot results in Fig.~\ref{fig:heuristics} only for the NN heuristic which achieves the smallest Bayes risk after the maximum time or the maximal number of measurements. There also exist slightly more sophisticated versions of the cross-entropy method, for instance using Gaussian distributions with an adaptive variance \cite{pourchot2018cem}. Note that the cross-entropy for \textit{discrete actions} is a reinforcement learning algorithm \cite{schuff2020improving}, while the cross-entropy method for \textit{continuous action spaces}, which we used in this work, is despite its similar name an evolutionary strategy.
\subsection{Trust region policy optimization}
We use TRPO \cite{schulman2015trust} with the  default multilayer perceptron policy as implemented in Stable Baselines 2.9.0 \cite{stable-baselines}. Hyperparameters which differ from their default values are $\gamma=1$, $\lambda=0.92$ and  $\textrm{vf}_\text{stepsize}=0.0044$. These hyperparameters are found to yield good results during initial testing of the algorithm, without doing a rigorous hyperparameter tuning. In particular, we do not use discounted rewards by setting the discount factor $\gamma=1$, because even without reward damping the rewards are defined such that they tend to decay exponentially with the number of experiments. Further, smaller values for $\lambda$ such as $\lambda=0.92$ often work well together with large $\gamma$. 

Pretraining is also implemented in Stable Baselines 2.9.0 \cite{stable-baselines} using the Adam optimizer \cite{kingma2014adam}. Deviations from the default parameters \cite{stable-baselines}  are the following: we use $10^4$ ($10^3$) episodes, sampled from a known heuristic, as an expert dataset. Pretraining runs with $10^4$ ($10^3$) epochs (the number of training iterations on the  expert dataset) and a batch size for the expert dataset of 100 ($10$) (bracketed values are used for the time-limited $(\omega, T_2^{-1})$ estimation to speed up the computation).

Training runs for 500 iterations. However, we use an exit condition which can stop the training earlier. The exit conditions stops training if the policy entropy drops below $0.005$ \cite{stable-baselines}.

TRPO is pretrained once per heuristics (PGH, $\sigma^{-1}$, the best of five CEM-trained NNs), and then trained $5$ times for each of the pretrainings, and we plot the results in Fig.~\ref{fig:heuristics} only for the best NN heuristic, i.e., with the largest Bayes risk at the end of the episodes.
\subsection{Resource limits of the RL environments}
Table \ref{tab:1} shows the limits on the number of experiments and the available time for all estimation problems. For numerical convenience, time-limited (experiment-limited) problems also have a 
limit on the number of measurements (the available time). In particular, for time-limited problems a limit on the number of experiments helps to avoid that a RL agent chooses very many small experiment times
which would lead to very long episodes, i.e., to very long run times. Also from a physical perspective, small experiment times are not sensible due to overheads for readout and preparation. Such overheads can be taken into account as a dead time of the sensor between experiments which can be easily implemented as a part of the RL environments. In case of experiment-limited environments, the time limit $10^{27}$ avoids issues with large numerical values and it is chosen large enough for the exp-sparse heuristic with 500 experiments, which chooses exponentially increasing times.

\onecolumngrid

\begin{table}[h]
	\begin{center}
		\begin{tabular}{ l r r }
			estimation problem & experiment limit & ~~~~~time limit \\
			\hline 
			$\omega$ estimation, $T_2=\infty$, experiment-limited & 20 & $10^{27}$ \\ 
			$\omega$ estimation, $T_2=\infty$, time-limited & 100 & 100 \\
			$\omega$ estimation, $T_2=100$, experiment-limited & 125 & $10^{27}$ \\ 
			$\omega$ estimation, $T_2=100$, time-limited & 1000 & 2500 \\ 
			$\omega$ estimation, $T_2=10$, experiment-limited & 125 & $10^{27}$ \\ 
			$\omega$ estimation, $T_2=10$, time-limited & 1000 & 2500 \\
			$(\omega, T_2^{-1})$ estimation, experiment-limited & 500 & $10^{27}$ \\ 
			$(\omega, T_2^{-1})$ estimation, time-limited & 4000 & 2500 \\  
		\end{tabular}
		\caption{Limits on the number of experiments and the available time for all estimation problems considered in this work.} \label{tab:1}
	\end{center}
\end{table}
\twocolumngrid
\section{Appendix E: Comparing experiment designs from different heuristics}
\begin{figure}[h]
	\centering
	\fbox{\includegraphics[width=0.48\linewidth]{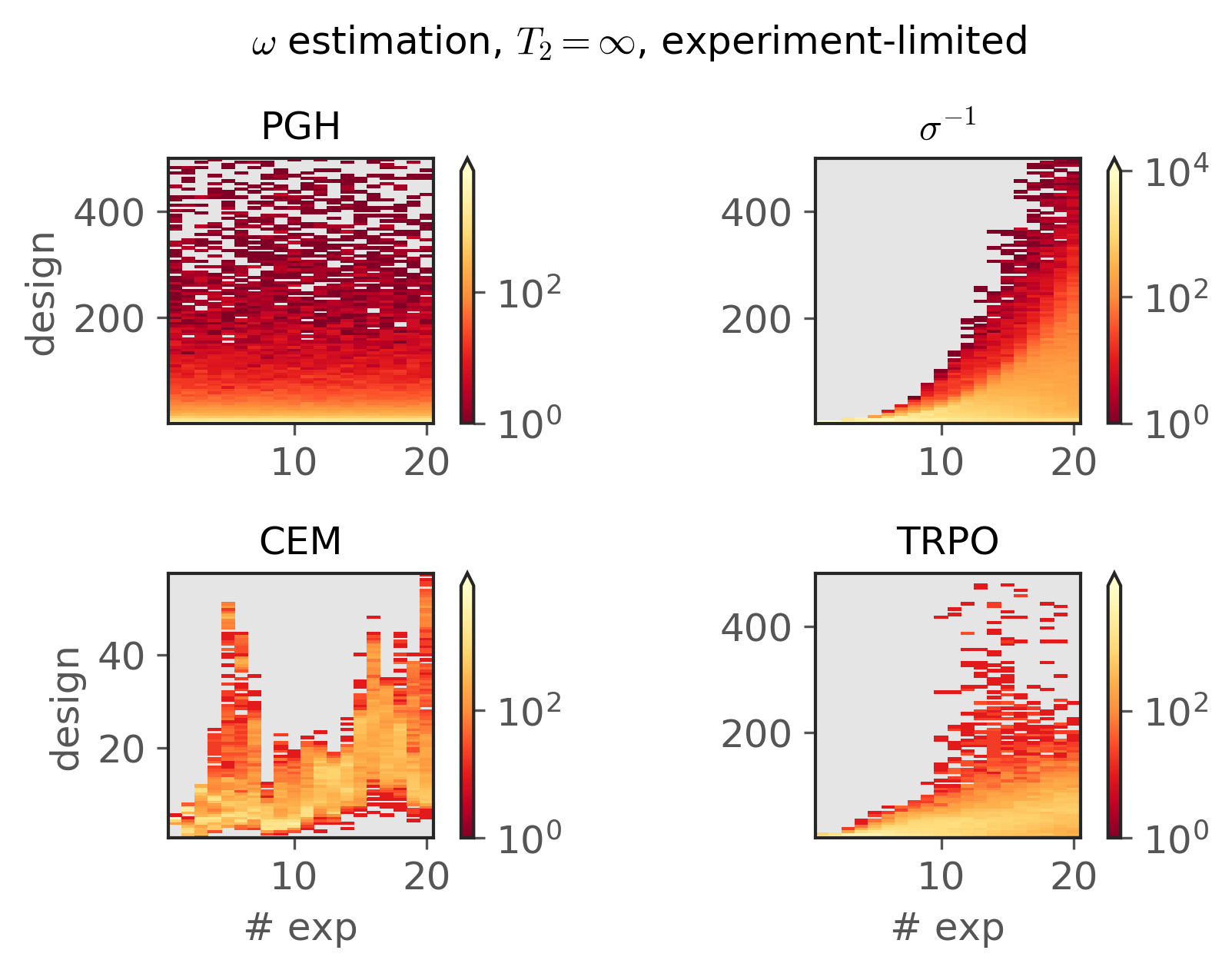}}\hfill
	\fbox{\includegraphics[width=0.48\linewidth]{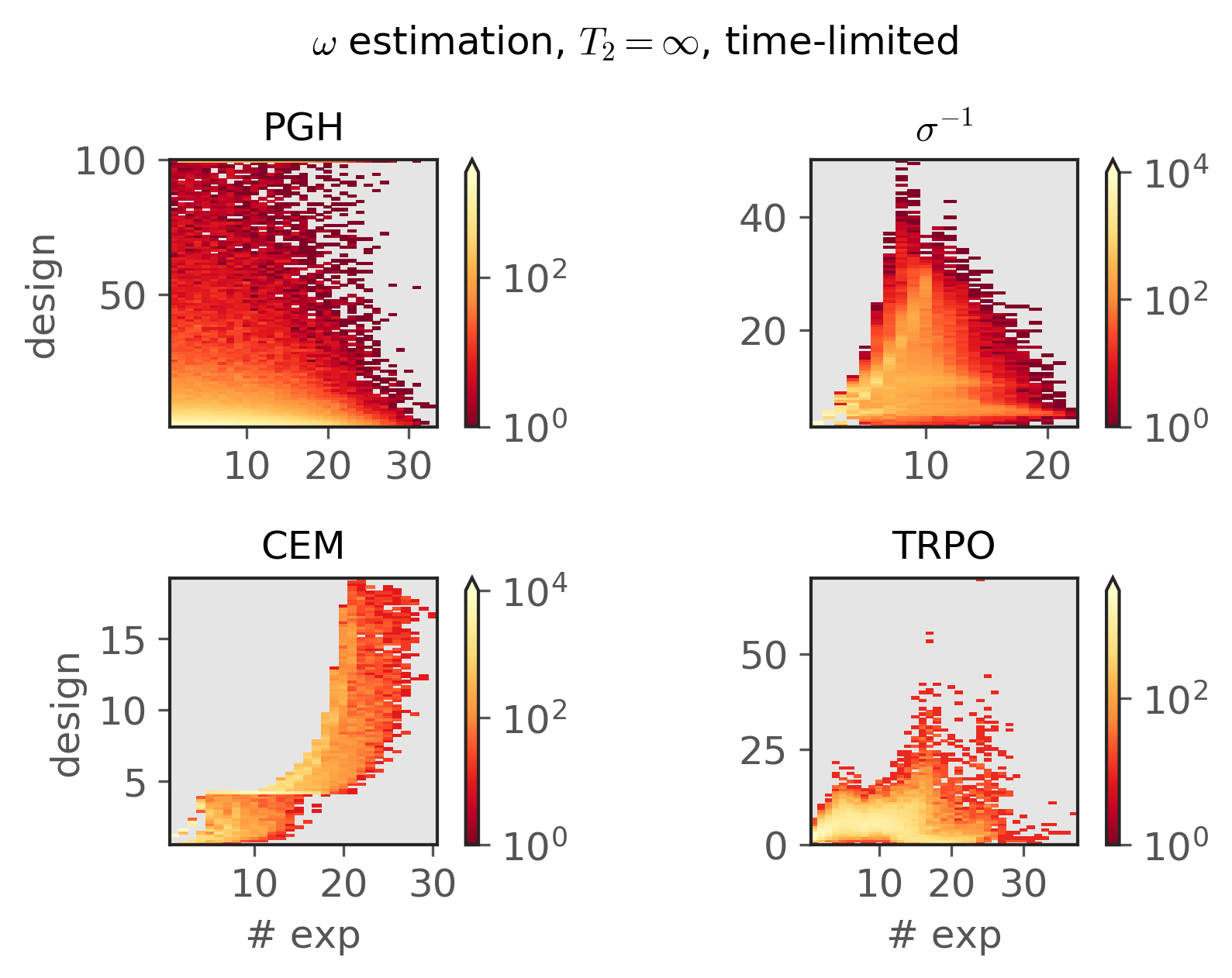}}\\
	\fbox{\includegraphics[width=0.48\linewidth]{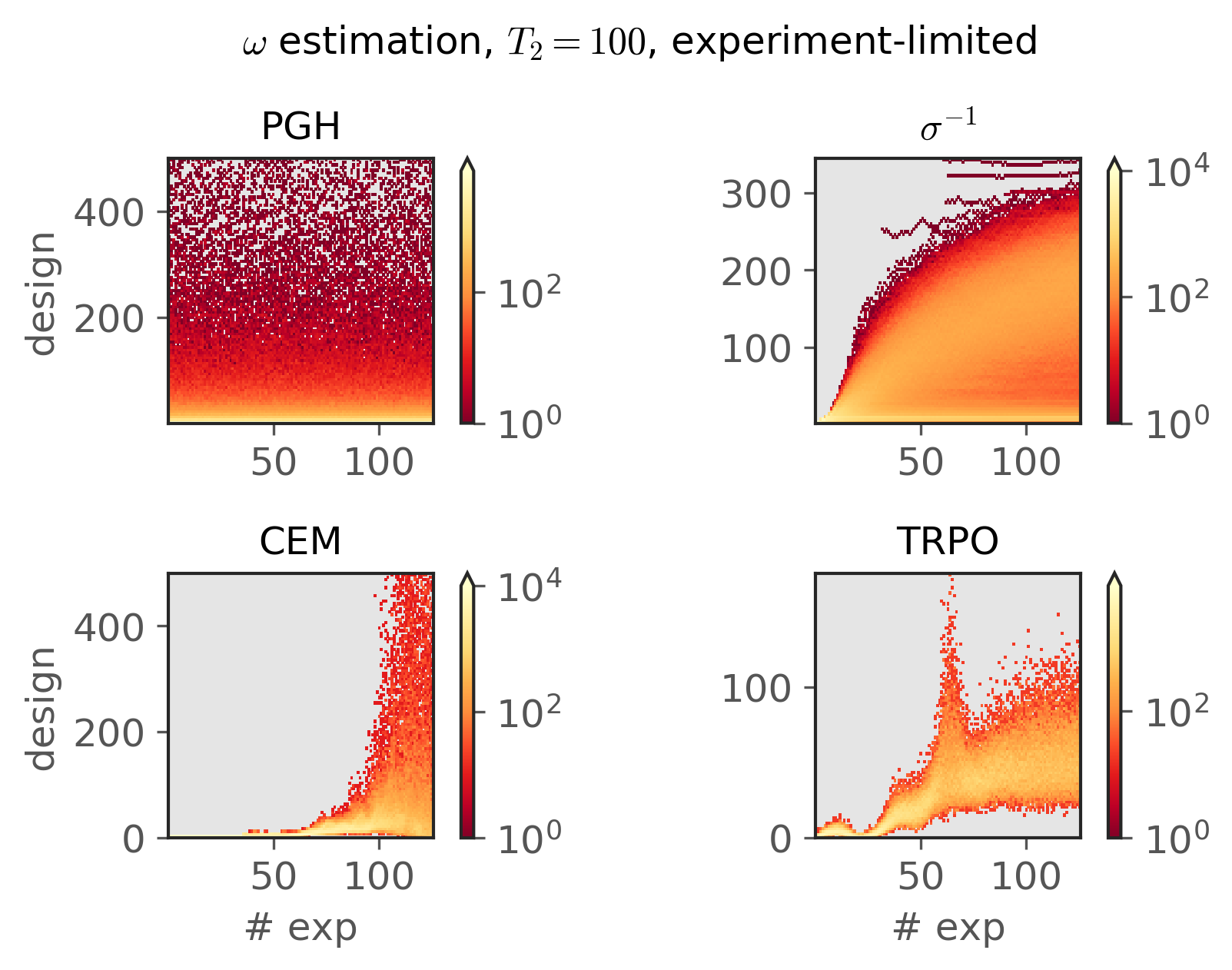}}\hfill
	\fbox{\includegraphics[width=0.48\linewidth]{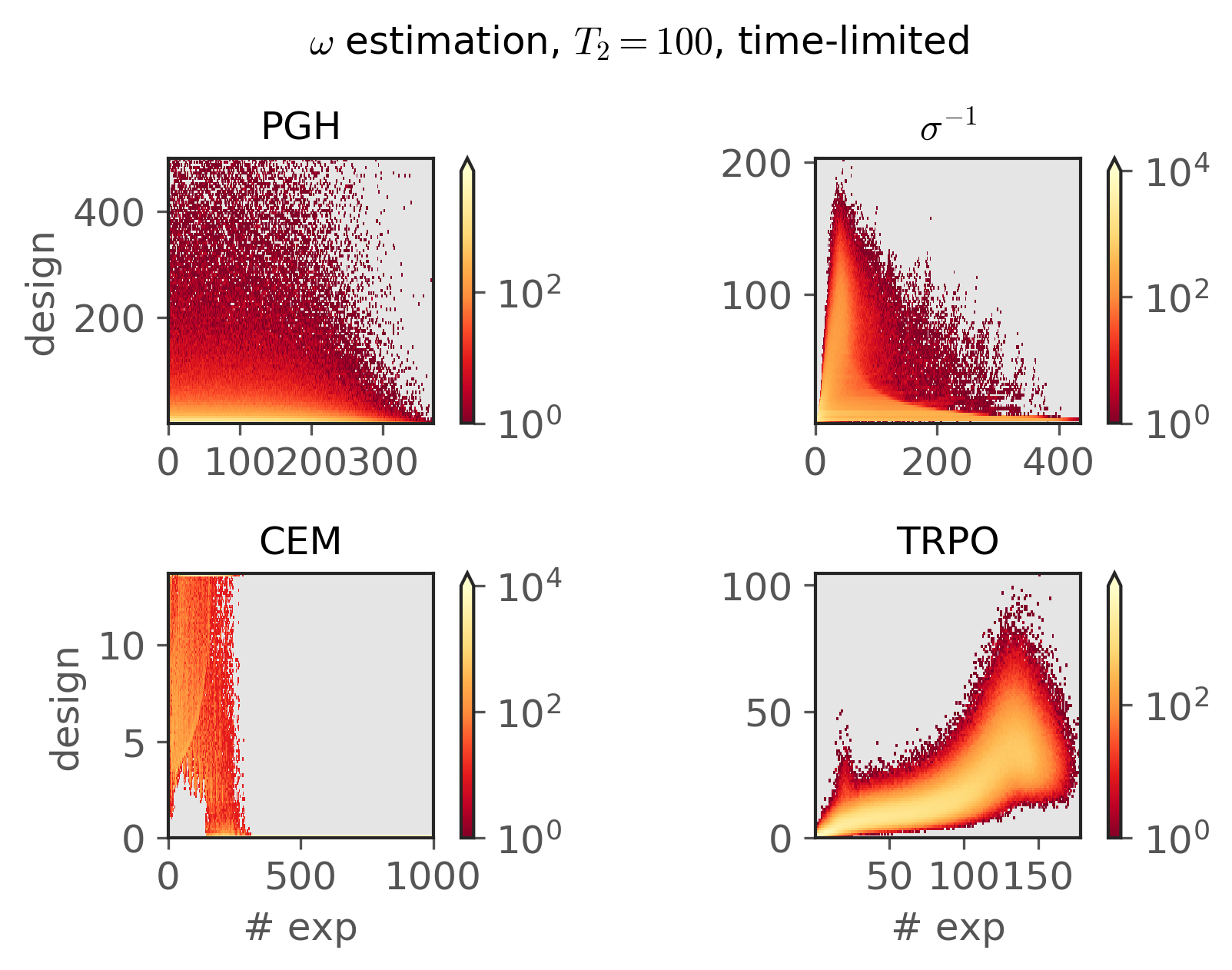}}
	\caption{Distribution of experiment designs ($y$ axis). On the $x$ axis we show the experiment number, e.g., $x=10$ corresponds to the $10$th experiment in each episode. For the sake of readability, we cut off $y$ data at $t=500$. Note that in the time-limited cases, CEM sometimes reaches the numerical threshold for the number of experiments by choosing many experiments with very small $t$. This manifests itself in the corresponding plots as a bright line close to $t=0$ (see for example the case of $\omega$ estimation, $T_2=100$, time-limited). If the gray background of the plots is visible, there are no experiment designs for the corresponding values. }\label{fig:sup1}
\end{figure}

\begin{figure}[h]
	\centering
	\fbox{\includegraphics[width=0.48\linewidth]{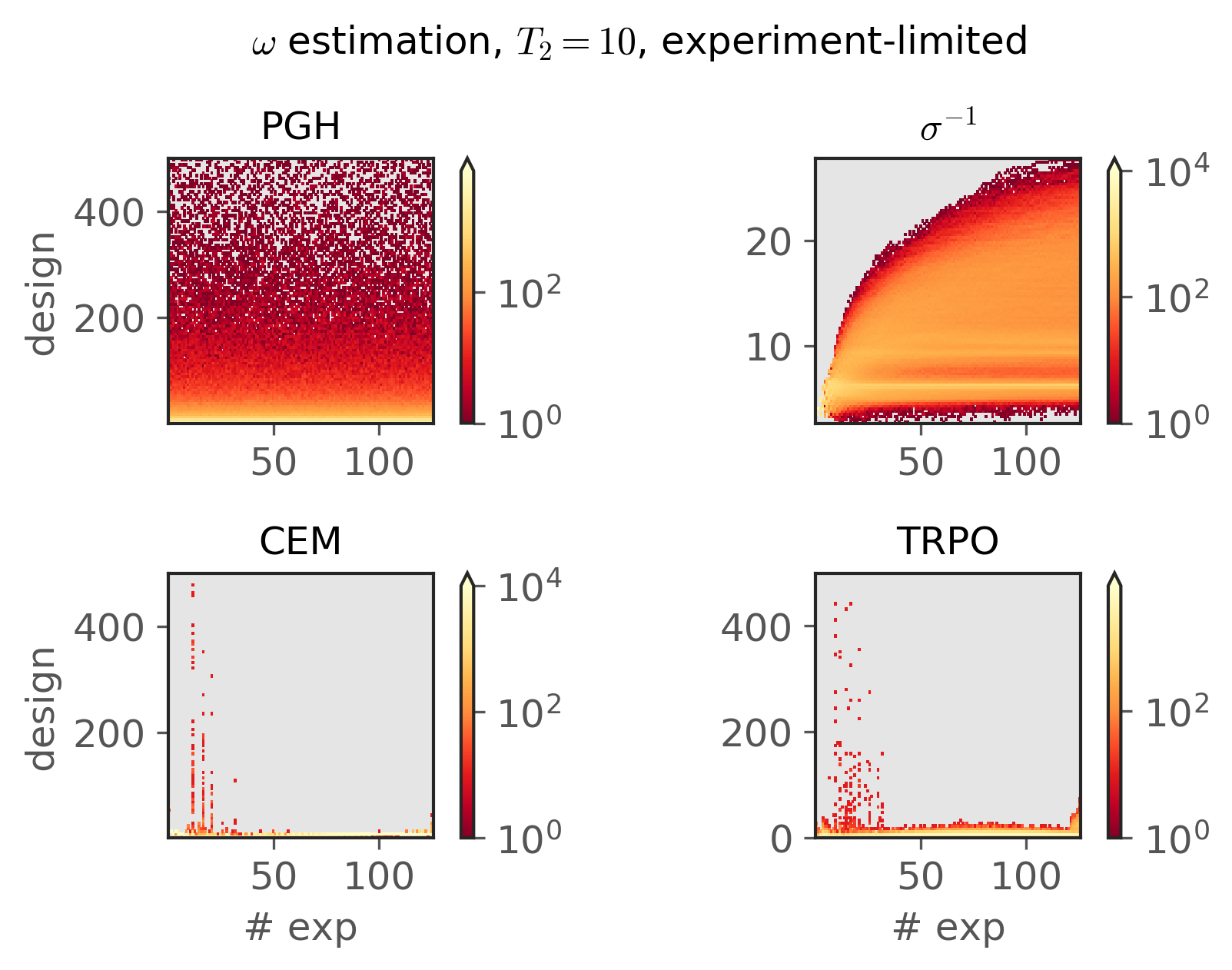}}\hfill
	\fbox{\includegraphics[width=0.48\linewidth]{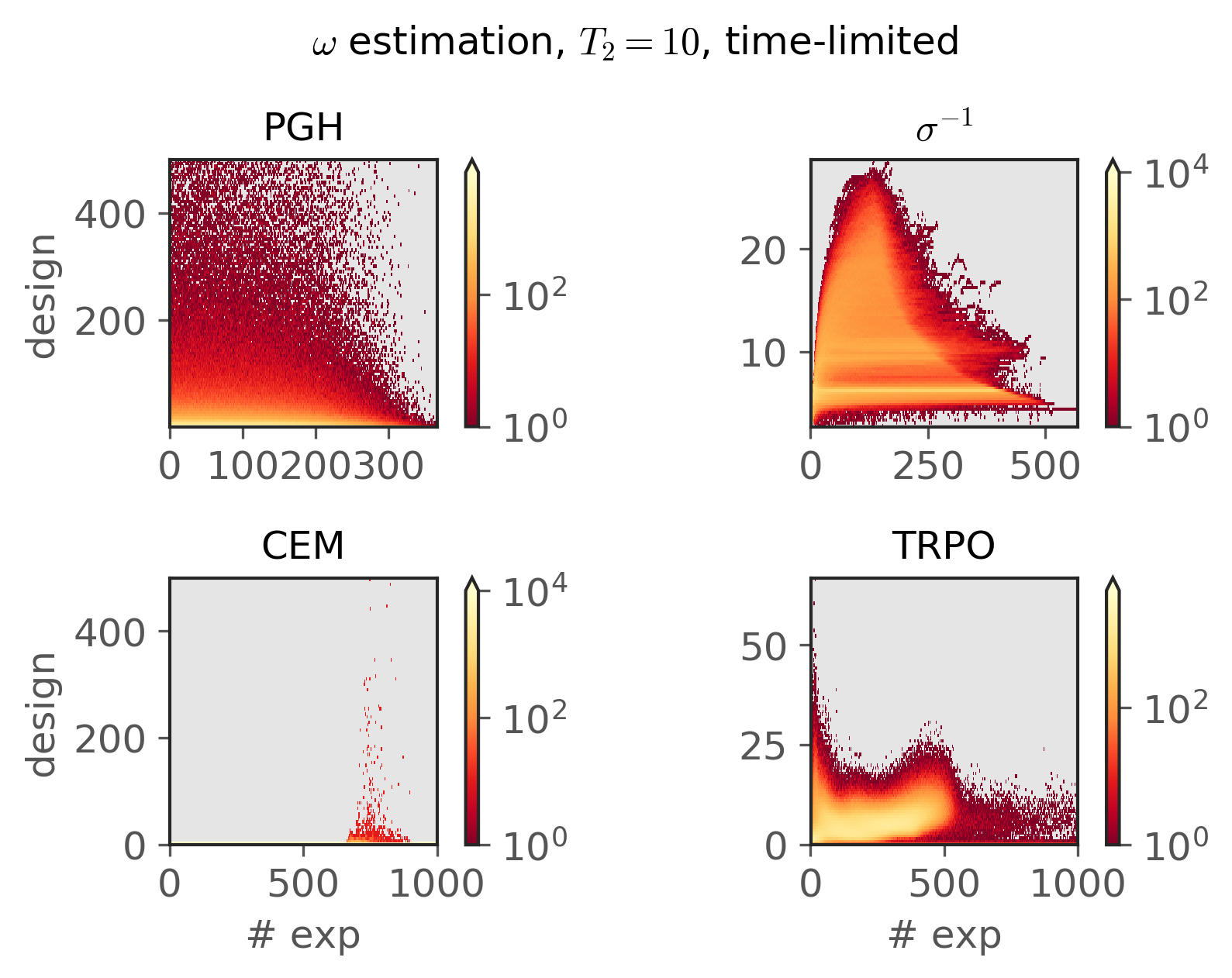}}
	\fbox{\includegraphics[width=0.48\linewidth]{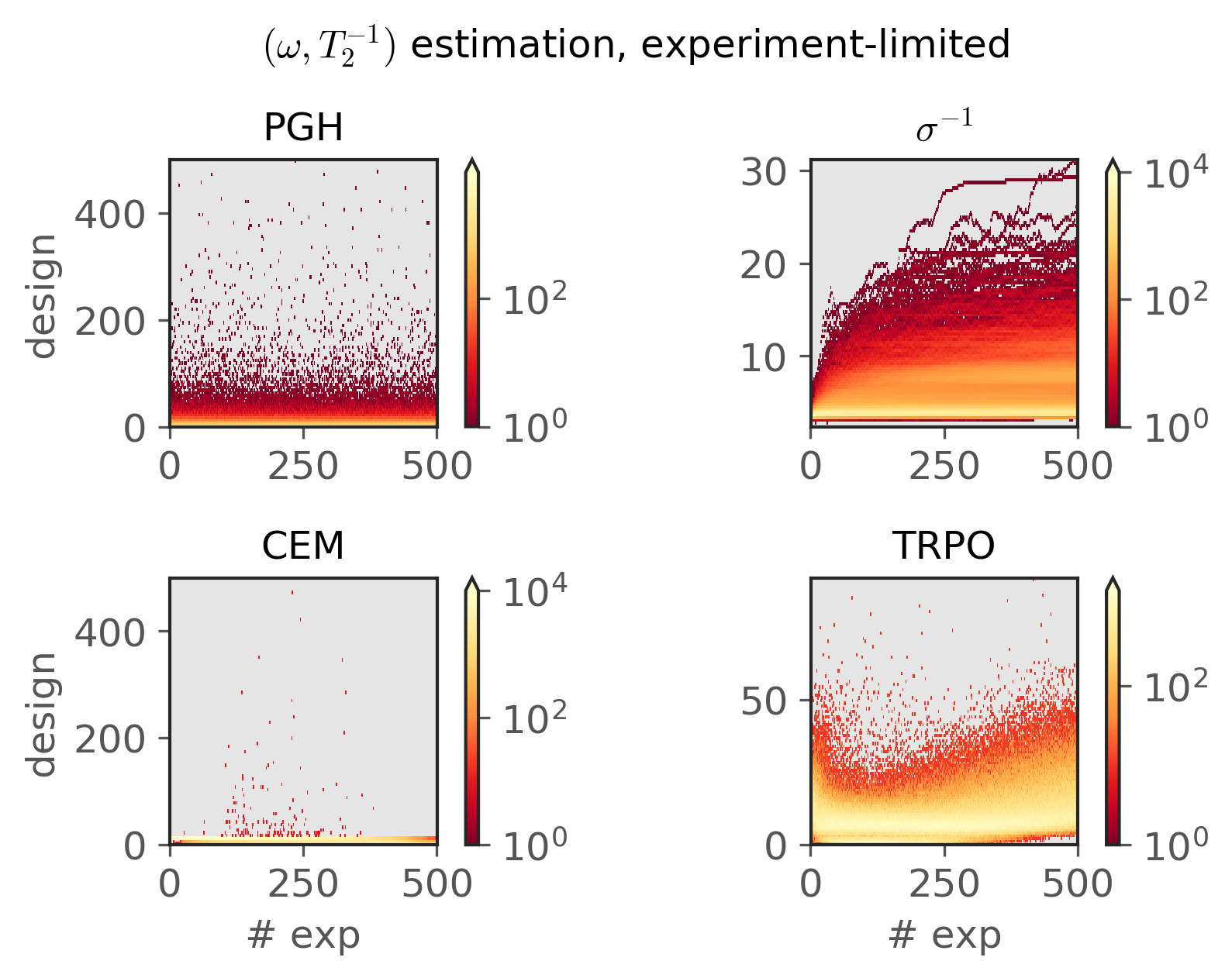}}\hfill
	\fbox{\includegraphics[width=0.48\linewidth]{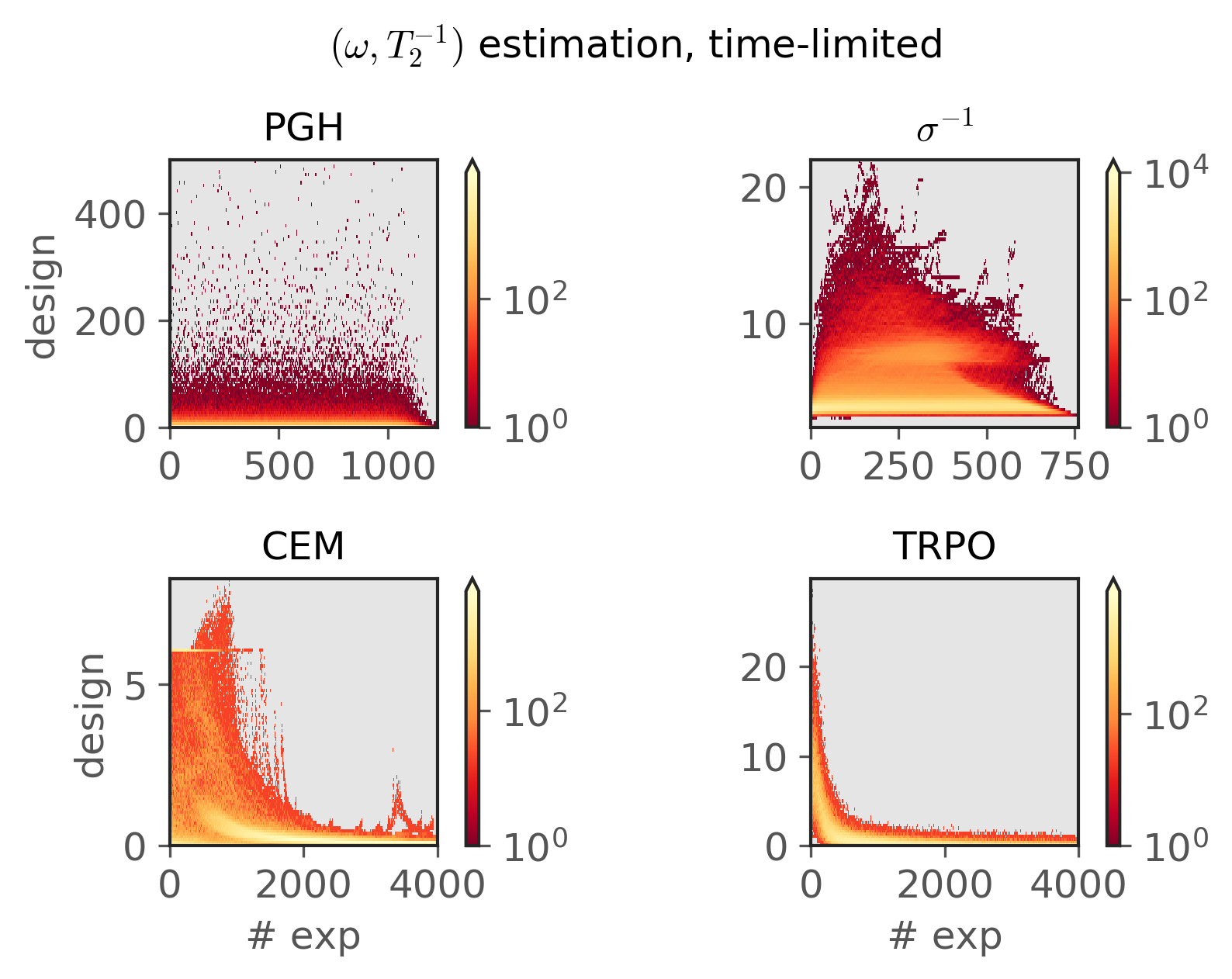}}
	\caption{See the caption of Fig.~\ref{fig:sup1}, the only difference is that this figure shows data for different estimation problems as specified in the titles of the subfigures).}\label{fig:sup2}
\end{figure}

Here we provide additional information about the distribution of experiment designs, i.e., experiment times, chosen by the four adaptive heuristics discussed in the main text: the particle-guess heuristic (PGH), the $\sigma^{-1}$ heuristic, a neural network trained with the cross-entropy method (CEM), and a neural network trained with trust region policy optimization (TRPO). Figs.~\ref{fig:sup1} and \ref{fig:sup2} show the frequency of experiment designs for different experiments, i.e., which times are chosen for the first experiment of each episode (Bayesian estimation), which for the second experiment, and so on. The plotted experiment designs correspond precisely to the data used for Fig.~\ref{fig:heuristics}. Data consist for each estimation problem and each heuristic of $10^4$ episodes. The diversity we find for different experiment-design heuristics is remarkable. Also the diversity between different environments underpins that experiment-design heuristics should be tuned to the particular estimation problem. However, the data presented in Figs.~\ref{fig:sup1} and \ref{fig:sup2} do not reveal information about the adaptivity of the heuristics.

\onecolumngrid

\clearpage

\end{document}